\documentclass[12pt,preprint]{aastex}

\begin{document}

\title{Grand Design and Flocculent Spirals in the $Spitzer$ Survey of Stellar
Structure in Galaxies (S$^4$G)}

\author{Debra Meloy Elmegreen\altaffilmark{1},
Bruce G. Elmegreen\altaffilmark{2},
Andrew Yau\altaffilmark{1},
E. Athanassoula\altaffilmark{3},
Albert Bosma\altaffilmark{3},
Ronald J. Buta\altaffilmark{4},
George Helou\altaffilmark{5,6},
Luis C. Ho\altaffilmark{7},
Dimitri A. Gadotti\altaffilmark{8,9},
Johan H. Knapen\altaffilmark{10,11},
Eija Laurikainen\altaffilmark{12,13},
Barry F. Madore\altaffilmark{7},
Karen L. Masters\altaffilmark{14},
Sharon E. Meidt\altaffilmark{15},
Kar\'{i}n Men\'{e}ndez-Delmestre\altaffilmark{7},
Michael W. Regan\altaffilmark{16},
Heikki Salo\altaffilmark{12},
Kartik Sheth\altaffilmark{17,5,6},
Dennis Zaritsky\altaffilmark{18},
Manuel Aravena\altaffilmark{17},
Ramin Skibba\altaffilmark{18}
Joannah L. Hinz\altaffilmark{18},
Jarkko Laine\altaffilmark{12,13},
Armando Gil de Paz\altaffilmark{19},
Juan-Carlos Mu\~noz-Mateos\altaffilmark{17,19},
Mark Seibert\altaffilmark{7},
Trisha Mizusawa\altaffilmark{17,5,6},
Taehyun Kim\altaffilmark{17,20},
Santiago Erroz Ferrer\altaffilmark{10,11}}
\altaffiltext{1}{Vassar College, Dept. of Physics and Astronomy, Poughkeepsie, NY 12604}
\altaffiltext{2}{IBM Research Division, T.J. Watson Research Center, Yorktown Hts., NY 10598}
\altaffiltext{3}{Laboratoire dÕAstrophysique de Marseille (LAM), UMR6110, Universit\'e de Provence/CNRS, Technop\^ole de Marseille Etoile, 38
rue Fr\'ed\'eric Joliot Curie, 13388 Marseille C\'edex 20, France}
\altaffiltext{4}{Department of Physics and Astronomy, University of Alabama, Box 870324, Tuscaloosa, AL 35487}
\altaffiltext{5}{Spitzer Science Center}
\altaffiltext{6}{California Institute of Technology, 1200 East California Boulevard, Pasadena, CA 91125}
\altaffiltext{7}{The Observatories, Carnegie Institution of Washington, 813 Santa Barbara Street, Pasadena, CA 91101}
\altaffiltext{8}{Max-Planck-Institut f\"ur Astrophysik, Karl-Schwarzschild-Strasse 1, D-85748 Garching bei Munchen, Germany}
\altaffiltext{9}{European Southern Observatory, Casilla 19001, Santiago 19, Chile}
\altaffiltext{10}{Instituto de Astrof\'\i sica de Canarias, E-38200 La Laguna, Spain}
\altaffiltext{11}{Departamento de Astrof\'\i sica, Universidad de La Laguna, E-38200 La Laguna, Spain}
\altaffiltext{12}{Department of Physical Sciences/Astronomy Division, University of Oulu, FIN-90014, Finland}
\altaffiltext{13}{Finnish Centre for Astronomy with ESO (FINCA), University of Turku}
\altaffiltext{14}{Institute of Cosmology and Gravitation, University of Portsmouth, Dennis Sciama Building, Burnaby Road,
Portsmouth, PO1 2EH, UK}
\altaffiltext{15}{Max-Planck-Institut f\"ur Astronomie, K¬onigstuhl 17, 69117 Heidelberg, Germany}
\altaffiltext{16}{Space Telescope Science Institute, 3700 San Martin Drive, Baltimore, MD 21218}
\altaffiltext{17}{National Radio Astronomy Observatory / NAASC, 520 Edgemont Road, Charlottesville, VA 22903}
\altaffiltext{18}{University of Arizona, 933 N. Cherry Ave., Tucson, AZ 85721}
\altaffiltext{19}{Departamento de Astrof\'\i sica, Universidad Complutense de Madrid, Madrid 28040, Spain}
\altaffiltext{20}{Astronomy Program, Dept. of Physics and Astronomy, Seoul National University, Seoul 151-742, Korea}
%\altaffiltext{23}{South African Astronomical Observatory, Observatory, 7935 Cape Town, South Africa}
%\altaffiltext{18}{Department of Physics and Astronomy, SUNY Stony Brook, Stony Brook, NY 11794-3800}
%\altaffiltext{20}{Department of Physics and Astronomy, University of California, Riverside, CA 92521}
%\altaffiltext{22}{NRC Herzberg Institute of Astrophysics, 5071 West Saanich Road, Victoria, V9E 2E7, Canada}
%\altaffiltext{11}{European Southern Observatory, Casilla 19001, Santiago 19, Chile}
\begin{abstract}
Spiral arm properties of 46 galaxies in the $Spitzer$
Survey of Stellar Structure in Galaxies (S$^4$G)
were measured at $3.6\mu$m, where extinction is small and the old stars dominate.
%The images include emission from old stars that trace spiral
%density waves, and also contain emission from red supergiants,
%which trace younger regions such as star complexes.
The sample includes flocculent, multiple arm, and grand design types
with a wide range of Hubble and bar types. We find that most optically
flocculent galaxies are also flocculent in the mid-IR because of star
formation uncorrelated with stellar density waves, whereas multiple arm
and grand design galaxies have underlying stellar waves. Arm-interarm
contrasts increase from flocculent to multiple arm to grand design
galaxies and with later Hubble types.
%with average arm-interarm
%contrasts of 0.3 to 1.5 mag, increasing with later Hubble types.
Structure can be traced
further out in the disk than in previous surveys. Some spirals peak at mid-radius while
others continuously rise or fall, depending on Hubble and bar type. We
find evidence for regular and symmetric modulations of the arm
strength in NGC 4321.
%as might be expected with wave interference, but we can only
%see concentrations of star formation at the peaks.
Bars tend to be
long, high amplitude, and flat-profiled in early type spirals, with arm
contrasts that decrease with radius beyond the end of the bar, and they
tend to be short, low amplitude, and exponential-profiled in late
Hubble types, with arm contrasts that are constant or increase with
radius. Longer bars tend to have larger amplitudes and stronger arms.
\end{abstract} 
\section{Introduction}

Measurements of spiral arm properties over a range of passbands provide
clues to the mechanisms of star formation and the generation of spiral
waves. The $Spitzer$ Space Telescope enables an extension to the
mid-IR, where extinction is small, light is dominated by old stars, and
the arms can be traced further out in the disk than in previous optical
and near-IR studies due to the greater sensitivity of the images. The
purpose of the present paper is to examine a representative sample of
galaxies with a range of spiral Arm Classes, Hubble types, and bar
types using the Spitzer Survey of Stellar Structure in Galaxies
(S$^4$G; Sheth et al. 2010).  Our motivations are to determine whether
the arm strengths in the mid-IR are similar to those in the optical and
near-IR, to trace spiral arms far out in the disk, and to examine how
the arm structures and amplitudes vary with radius, Hubble type, and
different bar properties.

Spiral arms and bars have been measured and interpreted theoretically
for more than 50 years.  They have many key properties that should be
reviewed before presenting the new $Spitzer$ data. In the next section,
we consider these properties and what they imply about the origins of
spiral structure. Our S$^4$G survey confirms most of the previous
observations and introduces new questions about radial profiles of
spiral arm strengths, spiral structures in the far outer regions, and
interactions between bars and spirals.  In Section \ref{sect:data} we
describe the dataset. In Section 4, we discuss spiral symmetry (Sect.
\ref{sect:sym}), arm-interarm contrasts (Sect. \ref{sect:morph}),
Fourier transform analyses (Sect. \ref{sect:FTs}), spiral arm amplitude
modulations (Sect. \ref{sect:mod}), smooth outer arms (Sect.
\ref{sect:outer}), sharp outer arm edges (Sect. \ref{sect:sharp}), and
bar properties (Sect. \ref{sect:bar}). Our conclusions are in Section
\ref{sect:conc}.

\section{Prior Observations of Galactic Spirals and Bars} \label{history}

The linear wave theory of \cite{lin64} suggested that spiral structure
is an instability in the disk of old stars, with relatively weak waves
made visible optically by triggered star formation \citep{roberts69}.
Streaming motions in HI \citep{vis80} confirmed the density wave
picture for the grand design spiral M81, and early photometry based on
$B, V$, and $O$-band photographic plates \citep{schweiz76} and $I$-band
plates (Elmegreen \& Elmegreen 1984; hereafter EE84) showed spiral
modulations of the underlying stellar disk.  These modulations were
stronger than expected from the Lin-Shu linear wave theory, so the role
of star formation and color gradients in delineating the arms
diminished \citep{ee86}.

In contrast, old stellar waves could not be seen in many galaxies with
short and patchy spirals (EE84), suggesting that these ``flocculent''
types are nearly pure star formation (e.g., Seiden \& Gerola 1979;
Elmegreen 1981; EE84). Multiple arm galaxies with inner 2-arm symmetry
and several long arms were seen to be an intermediate case: they have
underlying stellar spirals that are often regular in the inner parts
and irregular in the outer parts \citep{EE84,EE95}.

The irregularities in flocculent and multiple arm galaxies agree with
theoretical predictions of random gravitational instabilities in the
gas \citep{gold65} and old stars \citep{julian66, kalnajs71}. The
symmetric spirals in some grand design galaxies could be tidal features
\citep[e.g.][]{dobbs10}, and other symmetric spirals, along with the
inner symmetric parts of multiple arm galaxies, could be wave modes
reinforced by reflection or refraction in the central regions and
amplified at corotation \citep{bertin89,bertin00}. The swing amplifier
theory of \cite{toomre81} may apply to both the irregular arms and the
regular wave modes (Athanassoula 1992; see also Athanassoula, Bosma, \&
Papaioannou 1987). Early computer simulations of galactic spirals
almost always showed patchy and multiple arms \citep{sellwood84},
although steady 2-arm spirals were possible if the conditions were
right \citep{thom90}. Modern theories suggest that some spirals and
rings may be orbital features in a bar potential
\citep{romgoma,romgomb,lia09a,lia09b,athan10}.

Observations at $K_s$-band confirmed the large amplitudes of stellar
waves in grand design and multiple arm galaxies \citep{rix93, regan94,
rix95, knapen95, block04a}. Flocculent spirals rarely have underlying
stellar waves even in $K_s$-band. Weak 2-arm spirals were first found
in some flocculent galaxies by \cite{block94}, \cite{thornley96},
\cite{thornley97} and \cite{E99}, and discussed by \cite{block99} and
\cite{seigar03}. \cite{kend11} find that about half of the 31 galaxies
studied from the Spitzer Infrared Nearby Galaxies Survey
\citep[SINGS]{ken03} have underlying 2-arm structure, and all of the
optically grand design galaxies are grand design in the mid-IR as well.
They found a correlation between outer spiral arms and bars and between
spirals and companion galaxies. Previously, \cite{kn}, \cite{seigar2},
and many others observed that global spiral patterns are associated
with bars or companions.

The current interpretation of spiral structure suggests that grand
design spiral arms and the symmetric inner parts of multiple arm
galaxies are primarily spiral waves. The gas falls into this potential
well, or that formed by the manifold spirals, and shocks, making a dust
lane.  Star formation follows quickly in the dense molecular gas, and
the clouds and OB associations disperse downstream. Most of the
structures in flocculent galaxies and the outer irregular arms of
multiple arm galaxies are presumably from local instabilities in either
the gas alone for flocculents, or in the gas plus stars for long
irregular arms. The difference depends on the stability of the stellar
part of the disk in the two-fluid, gas+star system. \cite{dobbs10p}
discussed the possibility that some of the differences in spiral arm
types might be determined from azimuthal gradients of cluster ages.

Stellar bars are another important feature of galaxy disks and
presumably play a role in driving disk evolution and spiral structure.
Although the fraction of barred galaxies remains the same in the
near-IR and the optical \citep{MD07}, in some cases they may appear to
be much more prominent at the longer passbands
\citep{hack83,eskridge00}. The properties of stellar bars were
investigated in early photometric studies too (e.g., Elmegreen \&
Elmegreen 1985, hereafter EE85; Baumgart \& Peterson 1986; Ann 1986).
Two types of bars were found using $B$ and $I$ bands (EE85) and $K$
band \citep{regan97, lauri07}. In early Hubble types, bars tend to be
high amplitude, long, and flat in their surface brightness radial
profile, with a radially decreasing spiral arm amplitude beyond the
bar. In late Hubble types, bars tend to be low amplitude, short, and
exponential in profile, with rising spiral arm amplitudes beyond the
bar.

Angular momentum exchange between bars, disks and halos was studied by
\cite{debat00}, \cite{athan02}, \cite{athan03}, \cite{val03},
\cite{athan09}, and others. They all predicted that bars should grow
over time.  This prediction is consistent with $K_s$-band observations
by \cite{EE07} that longer and higher amplitude bars correlate with denser,
faster-evolving galaxies.

Bar torques were studied by Buta, Block, and collaborators (Buta \&
Block 2001; Block et al. 2001, 2002; Buta et al. 2004, 2006;
Laurikainen et al. 2004, 2007; see also Combes \& Sanders 1981). The
ratio of the maximum azimuthal force from a bar or spiral to the radial
force from all of the matter inside that radius is a measure of the bar
or spiral strength and is commonly called the torque parameter.
\cite{block02} found a wide distribution of total bar + spiral torque,
$Q_g$, in a large near-infrared sample of galaxies, and suggested that
most galaxies have torques that drive disk evolution. \cite{buta05}
found a smoothly declining distribution function of relative bar
torque, with no clear separation between strong and weak bars. They
also found a correlation between bar and spiral torques, and noted that
later Hubble types have larger relative bar torques because the bulges
are weaker.

\cite{combes93} and \cite{athan02} simulated flat and exponential bars
by varying the inner rotation curve or halo concentration. Flat bars
tend to occur in galaxies with sharply rising inner rotation curves,
and exponential bars occur in galaxies with slowly rising rotation
curves. These two cases are halo-dominated and disk-dominated in the
bar region, leading \cite{athan02} to call them types ``MH" and ``MD,"
respectively. Two bar types were also modeled by \cite{athan09} and
compared to observations by \cite{gadotti07}. They noted that flat bars
exchange more angular momentum with the halo than exponential bars.
This may explain why galaxies with flat bars tend to have 2-arm
spirals, while those with exponential bars have multiple arm or
flocculent spirals (EE85). Related to this is the observation that
higher amplitude bars tend to have stronger arms, as measured by
bar-interbar and arm-interarm contrasts (EE85; Ann \& Lee 1987;
Elmegreen et al. 2007;  Buta et al. 2009; Salo et al. 2010). This
correlation suggests that bars drive spirals through angular momentum
exchange with the disk.

Fourier transforms provide an objective indicator of spiral arm and bar
amplitudes, as first measured by \cite{iye82}, \cite{schloss84}, EE85,
and \cite{considere88}. In a recent study, \cite{durbala09} did Fourier
decompositions of Sloan Digital Sky Survey $i$-band images of $\sim100$
isolated barred and non-barred intermediate-type galaxies, and
confirmed the earlier studies that the lengths and amplitudes of bars
decrease from early to late Hubble types.

Several studies have noted correlations or weak correlations between
bar length and bar ellipticity (a proxy for bar strength), including
recent work by \cite{gad10} and \cite{com10} and references therein. A
recent Galaxy Zoo study of over 3000 Sloan Digital Sky Survey galaxies
also shows a correlation between bar length and ellipticity, and notes
that longer bars are associated with earlier type galaxies
\citep{hoyle}.

In the following S$^4$G survey of spirals and bars, many of the
features found in the optical and near-IR are reproduced with better
clarity here and with greater extension into the outer disk. Such
similarity between global structures over a wide range of wavelengths
suggests that these structures are massive. Because the $3.6\mu$m and
$4.6\mu$m images also contain hot dust \citep{rix95}, PAH emission
\citep{draine07}, AGB stars \citep{meidt11}, and red supergiants, there
is a resemblance among different passbands in the star-formation
features too. Future images with these contaminants removed
\citep{meidt11} could reveal other aspects of spirals and bars that are
not observed here.

\section{Data}

\label{sect:data} The S$^4$G is a survey of 2331 nearby galaxies in
$3.6\mu m$ and $4.5\mu m$ using channels 1 and 2 of the Infrared Array
Camera \cite[IRAC;][]{fazio04} of the Spitzer Space Telescope. It
includes archival and warm mission observations. The images reach a
limiting surface brightness of $\sim27$ mag arcsec$^{-2}$, which is
deeper than most previous spiral arm studies. The galaxies were all
processed or re-processed uniformly through the S$^4$G pipeline \citep{sheth10}. For the current study, we used the $3.6 \mu m$ images,
with a pixel scale of $0.75^{\prime\prime}$ and a resolution of
$1.7^{\prime\prime}$.  The $4.5\mu$m images were also examined, but the
results were generally similar to those at $3.6\mu$m so we do not
discuss them specifically. Both passbands highlight the old stellar
disks of spiral galaxies.

Our 46 galaxies were selected to include representative Hubble types Sa
through Sm, bar types SA, SAB, and SB, and Arm Classes flocculent,
multiple arm, and grand design. Half of our sample, 24 galaxies, is
part of the SINGS galaxy sample \citep{ken03}. In our sample, the
average blue-light radius $R_{25}$ (where the surface brightness is 25
mag arcsec$^{-2}$) is $3.6^{\prime}\pm2.4^{\prime}$, and the average
inclination is $41^\circ\pm13.2^\circ$ (from the Third Reference
Catalogue of Bright Galaxies, de Vaucouleurs et al. 1991, hereafter
RC3).  These galaxies constitute $\sim 20$\% of the S$^4$G sample with
$R_{25}>1.35^{\prime}$ and inclination $<60^\circ$. A more definitive
study of spiral and bar structures in galaxies will be possible when
the S$^4$G survey is complete.

Our sample is listed in Table \ref{table1} with spiral Arm Class, bar
type, and mid-IR Hubble type. Arm Classes are from the classification
system based on optical images \citep{EE87}, except for NGC 3049
classified here.  That paper has a 12-point subdivision of Arm Classes
to highlight details.  Here we combine the Arm Classes into the three
main divisions, so F (flocculent) refers to Arm Classes 1-3, M
(multiple arm) is Arm Classes 4-9, and G (grand design) is Arm Classes
10-12. Bar types (A=non-barred, AB=intermediate, B=barred) and mid-IR
Hubble spiral types (Sa through Sm) are from \cite{buta10}, except for
three which get their types from the RC3 (NGC 4450, 4689, and 5147).
The mid-IR types differ from the RC3 types by about 1 stage (earlier)
for intermediate type spirals, but are similar for earlier and later
types. What we call bar type and Hubble type here are synonymous with
``family classification'' and ``stage'' in \cite{buta10}.

Grand design, multiple arm, and flocculent types in the SA and SAB
category are illustrated in Figures \ref{GD}, \ref{MA}, and \ref{floc},
respectively; SB galaxies with a mixture of Arm Classes are in Figure
\ref{bar}. In each figure, the top row shows the $3.6\mu$m images
displayed on a logarithmic intensity scale. The second and third rows
have 2-fold and 3-fold symmetric parts enhanced (Sect. \ref{sect:sym}),
and the bottom row has the image reprojected in polar coordinates
(Sect. \ref{sect:morph}).

\section{Data Reduction and Analysis}
\label{reduc}
\subsection{Spiral Symmetry}\label{sect:sym}

In order to highlight disk symmetry, each image was rotated
$180^{\circ}$ using the IRAF task $rotate$,  and subtracted from the
original using $imarith$ to get the asymmetric part. Negative values
were set equal to 0 in the asymmetric images, using the IRAF task
$imreplace$, and this truncated image was then subtracted from the
original to get the symmetric part \citep[hereafter EEM]{EEM92}:
\begin{equation}
S_2(r,\theta)=I(r,\theta)-[I(r,\theta)-I(r,\theta+\pi)]_T.
\end{equation}
Here, $I$ is the original image, $S_2$ is the symmetric image, and $T$
stands for truncation. Sample 2-fold symmetric images are shown in the
second row of Figures \ref{GD}--\ref{bar}. The grand design symmetric
images look like cleaned-up images of the originals, with prominent
2-arm structure. It is difficult to remove stars globally from channel
1 images without also removing star-forming regions. These symmetric
images eliminate virtually all the stars and most of the star-forming
regions -- all but those that happen to be symmetric in the galaxy.
Typically 80-85\% of the light in these images is in the $S_2$ image
and 15-20\% is in the asymmetric image; most of this symmetric light is
from the azimuthally averaged disk, unrelated to spirals.

Three-arm structure can be highlighted by following a similar procedure
but rotating $120^{\circ}$ twice (EEM):
\begin{equation}
S_3(r,\theta)=2I(r,\theta)-[I(r,\theta)-I(r,\theta+2/3\pi)]_T -
[I(r,\theta)-I(r,\theta-2/3\pi)]_T
\end{equation}
The results are shown in the third row of Figures \ref{GD}--\ref{bar}.
The most prominent 3-arm structures appear in multiple arm galaxies. In
contrast, grand design galaxies are dominated by two arms. For the
flocculent galaxies, which do not have a dominant symmetric component,
the $S_3$ images resemble the $S_2$ images and both show a lot of arms.

Most of our galaxies have the same spiral Arm Class at all passbands.
The optically flocculent galaxies are also flocculent in $3.6\mu$m
because of patches from PAH emission and young supergiants. The
exceptions in our sample are the flocculents NGC 5055 and NGC 2841
(Fig. \ref{floc}), which have long and smooth spiral arms at $3.6\mu$m.
These arms were discovered by \cite{thornley96} in $K_s$-band and also
shown by \cite{buta10} and \cite{buta11}. The long-arm spiral in NGC
2841 is predominantly seen as a dust arm. NGC 7793 may have a subtle
2-arm component as well, which shows up weakly in the symmetric image.

To consider a broader sample of underlying long-arm structure in
flocculent galaxies, we examined 2MASS images of 197 galaxies
catalogued as optically flocculent by \cite{EE87}. We find that only
$\sim15$\% have underlying weak 2-arm structure in the near-IR 2MASS
images; most are still dominated by flocculent structure.

\subsection{Arm-Interarm Contrasts}
\label{sect:morph}

One measure of spiral arm amplitudes is a comparison of arm and
interarm surface brightnesses, which were measured on azimuthal scans
of deprojected images with sky subtracted. The position angles and
inclinations used for this are in Table \ref{table1}. Position angles
were measured from contours of the $3.6\mu$m images; if the outer and
inner disks had different orientations, then the position angle was
selected to be appropriate for the disk in the vicinity of the main
arms. The values we used are generally within a few degrees of those
listed in the RC3 or Hyperleda\footnote{http://leda.univ-lyon1.fr/}.
Slight differences like these changed the derived arm amplitudes by
less than $10\%$, because most of the galaxies in our sample have
relatively low inclination. \cite{gadotti07} discuss general errors for
measurements that depend on inclination, pitch angle, and the
assumption of circular disks.

Each galaxy image was transformed into polar coordinates, $(r,
\theta)$, as shown in the lower panels of Figures \ref{GD}--\ref{bar}. The y-axis is linear steps of radius, while the x-axis is the azimuthal angle from 0 to 360$^\circ$. From this polar image, $pvector$ was used to make
1-pixel-wide azimuthal cuts for radii in steps of 0.05$R_{25}$ (the
values we used for $R_{25}$, from RC3, are listed in
Table \ref{table1}).

Spiral arms were measured from the azimuthal scans, often out to
$\sim1.5R_{25}$, which is further than in previous near-IR studies
because of the greater sensitivity of the S$^4$G survey. Star-forming
spikes and foreground stars were obvious on the azimuthal scans and
avoided in the arm measurements. This avoidance of point sources gives
the arm-interarm contrast an advantage over Fourier transform
measurements in uncleaned images.

The arm-interarm intensity contrast at radius $r$ was converted to a
magnitude using the equation
\begin{equation}
A(r)=2.5\log\left[\frac{2I_{\rm arm}(r)}{I_{\rm interarm1}(r)+I_{\rm
interarm2}(r)}\right];
\end{equation}
$I_{arm}$ is the average arm intensity measured at the peak of the broad component,  and the denominator contains the
adjacent interarm regions. Figure \ref{armint} shows $A(r)$ in the top
row and in the third row from the top. Two galaxies from each of
Figures \ref{GD}--\ref{bar} are included. The average arm-interarm
contrast for each galaxy, averaged over the whole disk (beyond the bar,
if there is one), is listed in Table \ref{table2}; the errors are
$\sim$0.1 mag.

The average arm-interarm contrast in the disk versus Hubble type is
shown in Figure \ref{armHTAC}. Different Arm Classes have different
symbols. For a given Hubble type, the average arm-interarm contrast
decreases from grand design galaxies to multiple arm to flocculent.
This is consistent with the more prominent appearance of arms in grand
design spirals. The averages for all Hubble types combined are
$1.14\pm0.44$, $0.81\pm0.28$, and $0.75\pm0.35$ in these three Arm
Classes, respectively. Within a given Arm Class, the arm-interarm
contrast increases slightly for later Hubble types, probably because of
more star formation contributing to the arms; the later type galaxies
in our sample are all flocculent or multiple arm.

Sixteen galaxies in our sample were previously measured in optical and
near-IR bands. The arm-interarm contrasts at 3.6$\mu$m, and for some of
them at 4.5$\mu m$, are listed for the same radius as the optical and
near-IR results for these galaxies in Table \ref{table3}. The contrasts
are qualitatively similar between optical, near-IR, and mid-IR bands;
the average arm-interarm contrasts are 0.93$\pm$0.32, 0.81$\pm$0.49,
and 0.99$\pm$0.46 mag for $B$, $I$, and $3.6\mu$m, respectively, for all galaxies. For
multiple arm and grand design galaxies, the averages are
1.04$\pm$0.28, 0.99$\pm$0.46, and 1.17$\pm$0.38, which are all the same
to within the errors. The basic reason these spiral arm amplitudes are nearly independent of color is that the arms are intrinsically strong mass perturbations; color variations are relatively minor.

For flocculent galaxies, the contrasts are smaller than the multiple
arm and grand design galaxies and significantly larger in the blue than
at longer wavelengths: 0.63$\pm$0.24, 0.34$\pm$0.18, and 0.44$\pm$0.13
for for $B$, $I$, and $3.6\mu$m, respectively. This decrease in
amplitude from $B$ to $I$ band is most likely the result of a young age
for these spiral arm features. They appear to be star formation
superposed on a somewhat uniform old stellar disk. The slight increase
from $I$ band to $3.6\mu$m could result from PAH emission and red
supergiants in the $3.6\mu$m band. We attempted to avoid small regions
of emission in our measurements at $3.6\mu$m, so if there is
contamination from PAHs, it would have to be somewhat extended.
\cite{meidt11} are exploring techniques to remove PAH and point source
emission from the $3.6\mu$m images to produce mass maps.

Another explanation for decreasing arm-interarm contrast with
increasing wavelength in flocculents might be an excess of gas and dust
between the arms. The dust would depress the interarm brightness for
the $B$ band but not for $3.6\mu$m. While it is more likely that the
gas and dust are associated with star formation, which is what we see
as flocculent arms, it is still possible that shells and other debris
around the star formation site darken the interarm regions at short
wavelengths.

\subsection{Fourier Transforms}
\label{sect:FTs} Fourier transforms are an independent method for
measuring arm amplitudes besides arm-interarm contrasts, so Fourier
components of azimuthal intensity profiles on the $3.6\mu$m images were
also measured. For number of arms $m=2$, 3 and 4, the Fourier transform
was determined from the equation
\begin{equation}
F_{m}(r) = \frac{\sqrt{[\Sigma I(r,\theta) sin(m\theta)]^2 + [\Sigma
I(r,\theta) cos(m\theta)]^2}}{\Sigma I(r,\theta)}
\label{eqFT}.\end{equation} $I(r,\theta)$ is the intensity at azimuthal
angle $\theta$ and radius $r$ in the deprojected, sky-subtracted image.
All sums are over azimuthal angles with steps of 1 pixel. Note that
according to this definition, the relative amplitude of a spiral arm is
twice the value of the Fourier component. For example, a spiral arm
with an amplitude profile $ I(\theta)=1+A\sin(m\theta)$ has a Fourier
$m$-component from equation \ref{eqFT} equal to $A/2$.

Figure \ref{armint} shows Fourier components in the second and fourth
rows, below the corresponding arm-interarm contrasts.  (Fourier
transforms on the $4.5\mu$m images were indistinguishable from those at
$3.6\mu$m and are not considered further here).  The Fourier transforms
were done on images that were cleaned to remove most foreground stars;
masks for this process are part of Pipeline IV in the S$^4$G data
reduction. Some scans (such as NGC 5457) still contained some
foreground stars, and the Fourier components of these stand out as
narrow spikes. The star peaks are avoided in the average values used in
the discussion below.

Table \ref{table2} lists the average values of the $m=2$ components.
Uncertainties in the m=2 components range from $\sim\pm0.02$ to 0.10 in
different galaxies. The $m=4$ components are about half the $m=2$
components in all cases. The ratio of the $m=3$ component to the $m=2$
is largest for the flocculent galaxies, as expected for galaxies with
irregular structure and lots of arm pieces. Grand design galaxies are
dominated by two main arms, so their $m=3$ component is weaker. The
average values of $F_3$/$F_2$ are $0.58\pm0.11$, $0.48\pm0.24$, and
$0.33\pm0.19$ for flocculent, multiple arm, and grand design galaxies,
respectively.

In Figure \ref{fit}, the average $m=2$ value of the Fourier component
in the spiral region is compared with the average arm-interarm
contrast. The two measures of arm amplitude are obviously related. The
curves in the figure show the expected trends in two cases: the dashed
line follows if the arm-interarm contrast is given exclusively by the
amplitude of the $m=2$ Fourier component, and the solid line follows if
the arm-interarm contrast comes from the combined $m=2$ and $m=4$
components. If only the $m=2$ component contributes, then
$A/I=(1+2F_{2})/(1-2F_{2})$. If the $m=4$ component contributes half
the 2-arm amplitude of the $m=2$ component as found above, then
$A/I=(1+3F_{2})/(1-3F_{2})$. In the figure, the curves are
$2.5\log(A/I)$. The solid line traces the data points reasonably well,
but neither line is a good fit in a statistical sense because the
scatter in the data is large (the reduced chi-squared values comparing
the lines with the data are 0.69 for the solid line and 2.25 for the
dashed line, with $r$-values of 0.588 and 0.629, respectively).

Other studies have also measured peak $m=2$ Fourier amplitudes. For
comparison, \cite{lauri04} has 8 galaxies in common with ours from the
OSUBSGS; their $H$-band images yield amplitudes that are similar to the
$3.6\mu$m values, taking the factor of two (mentioned above) into
account to compare our $F_2$ values with their amplitudes.

\subsection{Spiral Arm Modulations in Radius}
\label{sect:mod}

In previous optical studies of symmetry images \citep{ESE89,EEM92}, we
saw large-scale amplitude modulations in spiral arms that looked like
interference between inward and outward moving waves in a spiral wave
mode \citep{bertin89}.  Evidence for these can be seen in the mid-IR
$m=2$ symmetric image of NGC 4321 in Figure \ref{GD}. The symmetric
image of NGC 4321 is shown deprojected with circles in Figure
\ref{sym4321}, along with arm-interarm contrast measurements and
Fourier transform measurements discussed in the previous sections, now
plotted with radius on a logarithmic scale. Referring to the top left
panel of Figure \ref{sym4321}, the parts of the arms at the ends of the
central oval (inner circle, at $\sim0.25R_{25}$) are more prominent
than the parts along the oval minor axis.  Further out in radius the
arms get brighter again (middle circle, at $\sim0.45R_{25}$); then
there is another gap and another brightening near the ends of the arms.
The outer circle in the figure is at $R_{25}$. These bright regions
trace the prominent trailing arms which should have an inward group
velocity \citep{toomre69}. If there are also leading spirals from a
reflection of the trailing spirals in the central parts, then the group
velocity of the leading spirals would be outward. The bright regions
could then be regions of constructive interference where these two
waves types intersect.

The lower left panel of Figure \ref{sym4321} shows the arm-interarm
contrast for each arm as in Figure \ref{GD}. The arrows at the bottom
are the radii for the arm maxima found before in $B$ and $I$-band images
\citep{ESE89}.  Both arms follow an alternating pattern of brightness
with a logarithmic spacing, although arm 1 is more variable than arm 2.
The arm-interarm contrast profile from the deprojected $m=2$ symmetric
image is shown in the lower right.  A modulation with the same log
spacing is present. The
two inner circles in the top left panel are at the radii of the two
peaks in the lower right panel. The strength of the arm modulation in
the symmetric image is 50\%-100\%.
Fourier transform amplitudes are shown in the top
right panel. The amplitude variations are most pronounced for the $m=4$
Fourier component, but they also show up in the $m=2$ component.

Close inspection of the $3.6\mu$m image shows that the inner peak
amplitude at $\sim0.25R_{25}$ is at the end of the inner oval, in a
broad region of bright star formation.  The next peak at
$\sim0.45R_{25}$ is the broad ridge of star formation in the main
spiral arm, presumably from density wave compression in a shock. The
outer peak in the arms is another star formation feature.  Because of
the mixture of old stars, PAH emission, and AGB stars in the $3.6\mu$m
image, it is difficult to tell if these peaks are present in the old
stellar component.  At the very least, it appears that the amplitude
modulations found here are the locations of extended star formation
regions.  Perhaps such star formation highlights an underlying wave
interference or resonance pattern \citep{knapen92}.

\subsection{Outer Spiral Arms}
\label{sect:outer}

\begin{figure}\epsscale{1}
\plotone{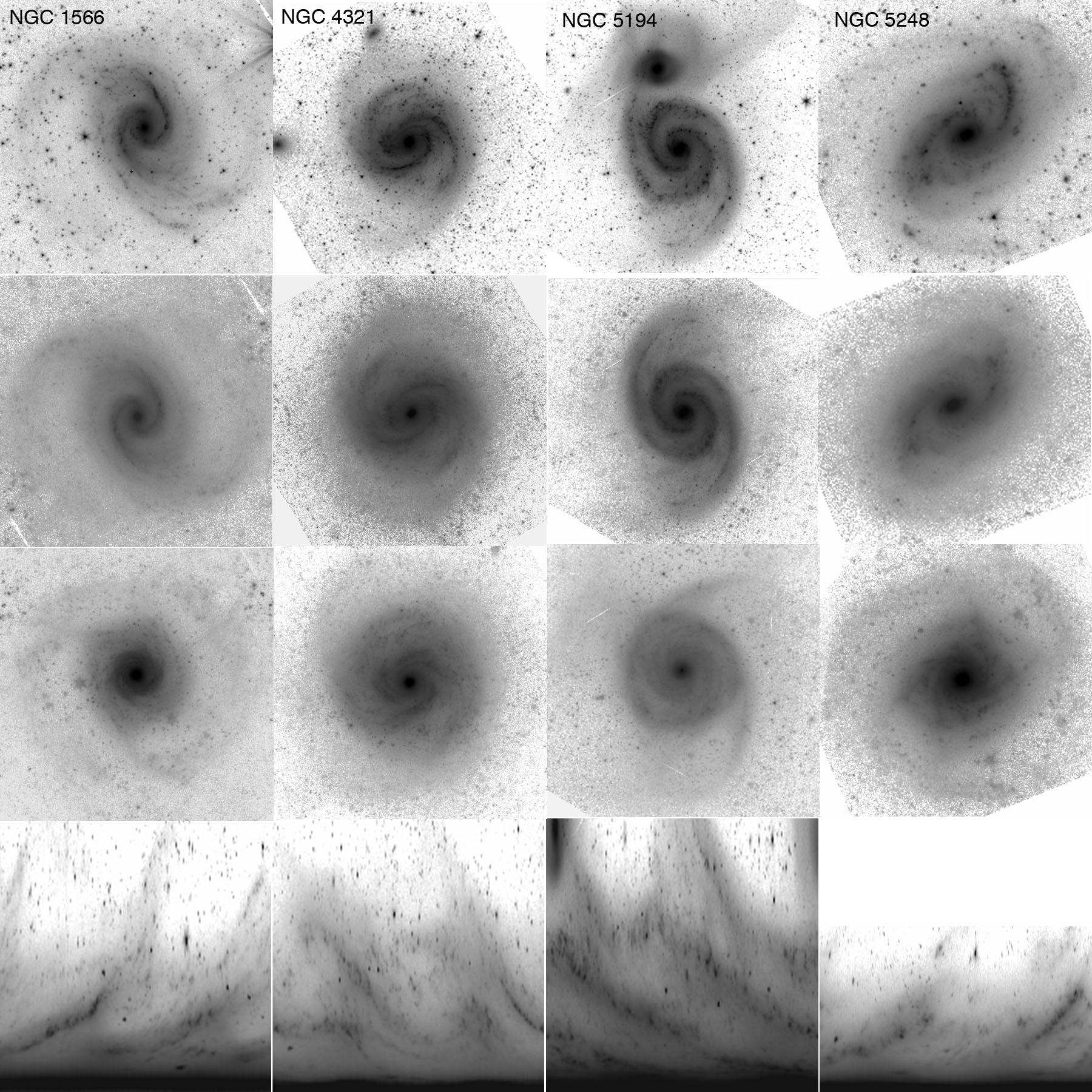}
%\plotone{arm-Fig1-1-2-11.jpg}
\caption{Top: Examples of grand design galaxies at 3.6$\mu$m shown on a
logarithmic intensity scale in sky-view, not deprojected. Top middle:
$S_2$ images, symmetric over $180^{\circ}$ as described in Section
\ref{sect:sym}. Lower middle: $S_3$ images, symmetric over
$120^{\circ}$. Bottom: Polar plots (R, $\theta$) showing radius on the
y-axis and azimuthal angle on the x-axis, based on deprojected images.
The lower part of the polar plot corresponds to the galaxy center; the
angle (x-axis) spans 0 to $360^{\circ}$, with arbitrary starting point
so that arms are not split. The outermost radius varies in each figure,
from about 1 to 1.5 $R_{25}$. }\label{GD}\end{figure}

\begin{figure}\epsscale{1}
\plotone{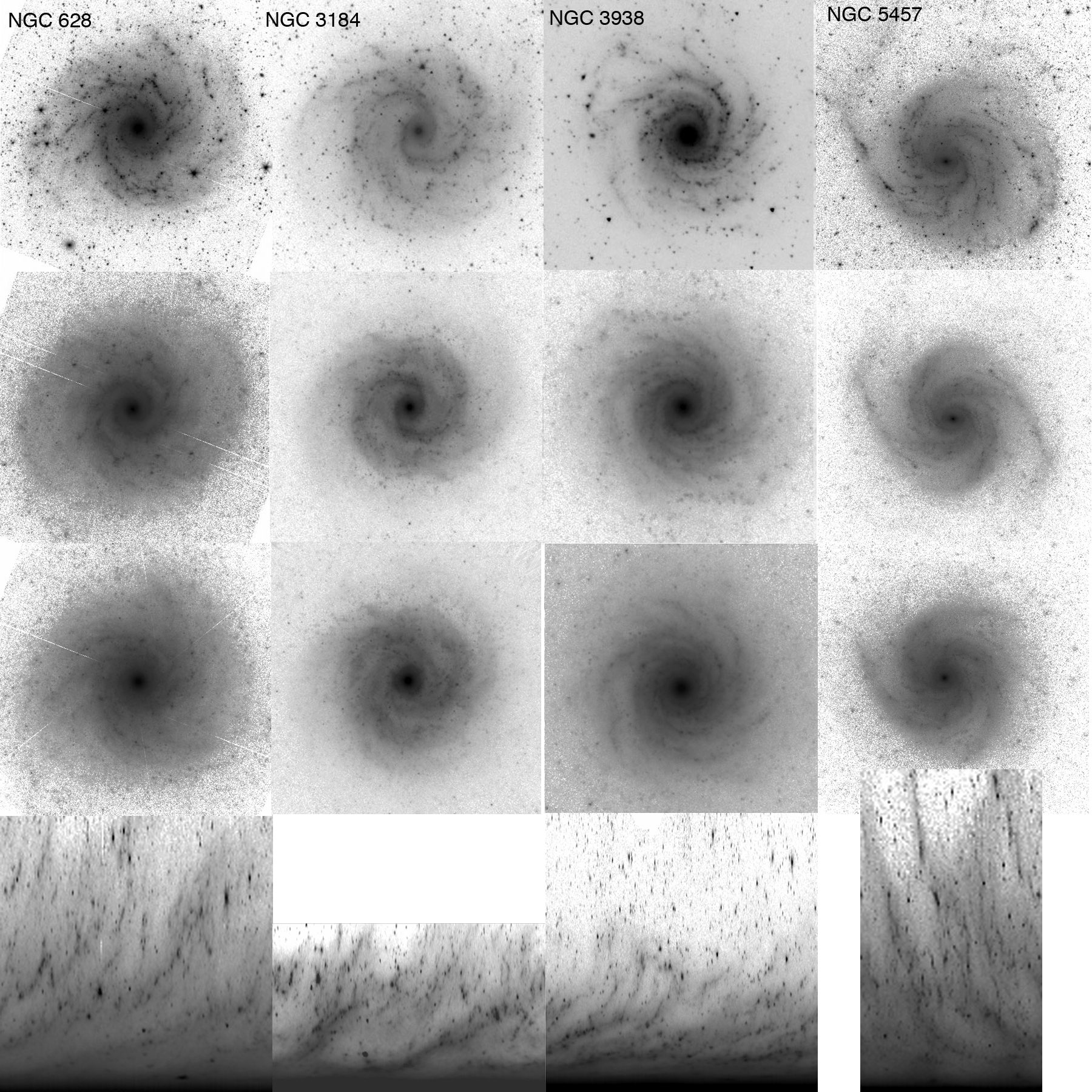}
%\plotone{arm-fig2-1-2-11.jpg}
\caption{Examples of multiple arm galaxies; same convention as in Figure \ref{GD}, with 3.6$\mu$m images in the top row, symmetric images in the middle (top middle: $S_2$; bottom middle: $S_3$) and polar plots on the bottom.}\label{MA}\end{figure}

\begin{figure}\epsscale{1}
\plotone{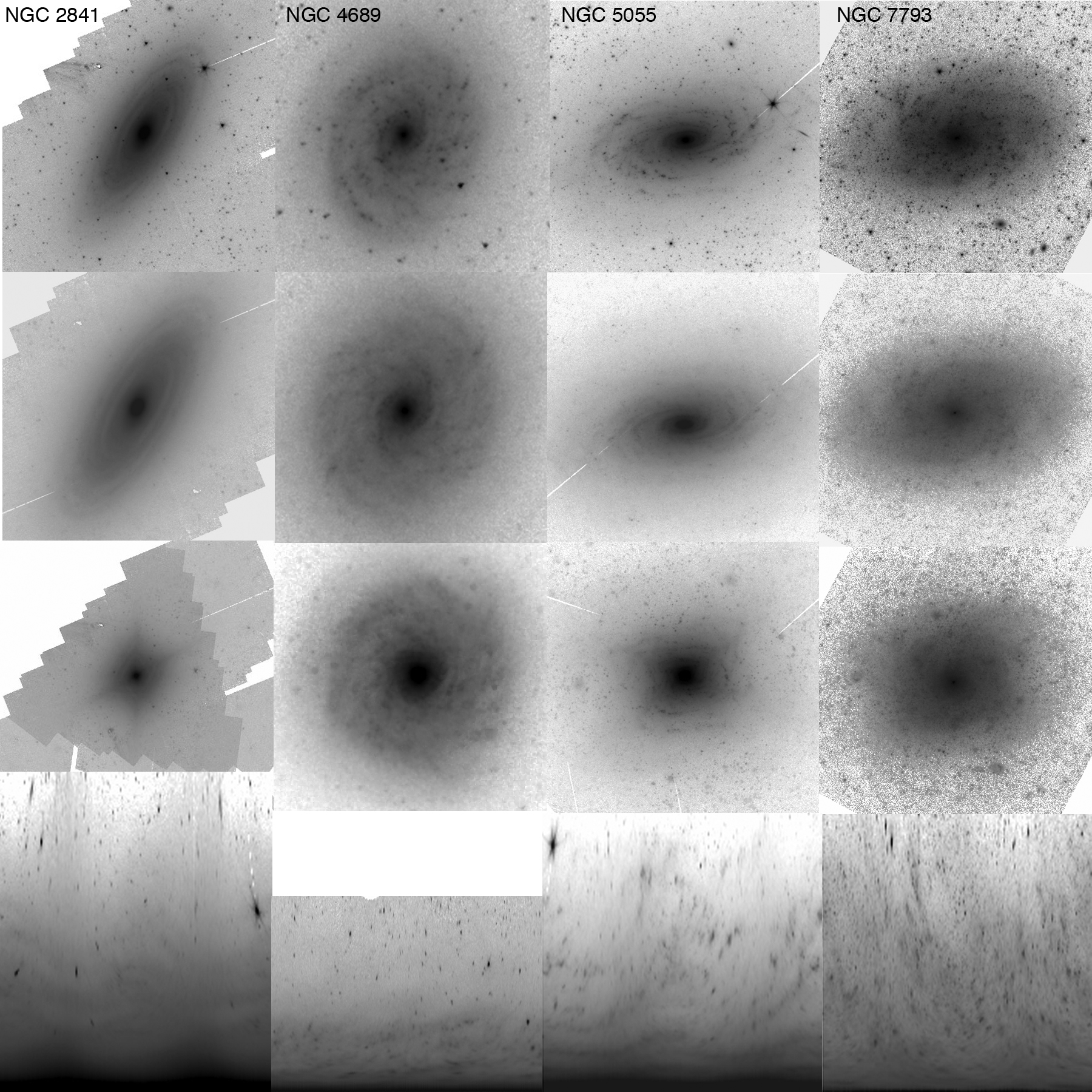}
%\plotone{arm-fig3-1-2-11.jpg}
\caption{Examples of flocculent galaxies; same convention as in Figure \ref{GD}, with 3.6$\mu$m images in the top row, symmetric images in the middle (top middle: $S_2$ images; bottom middle: $S_3$ images) and polar plots on the bottom.}\label{floc}\end{figure}

\begin{figure}\epsscale{1}
\plotone{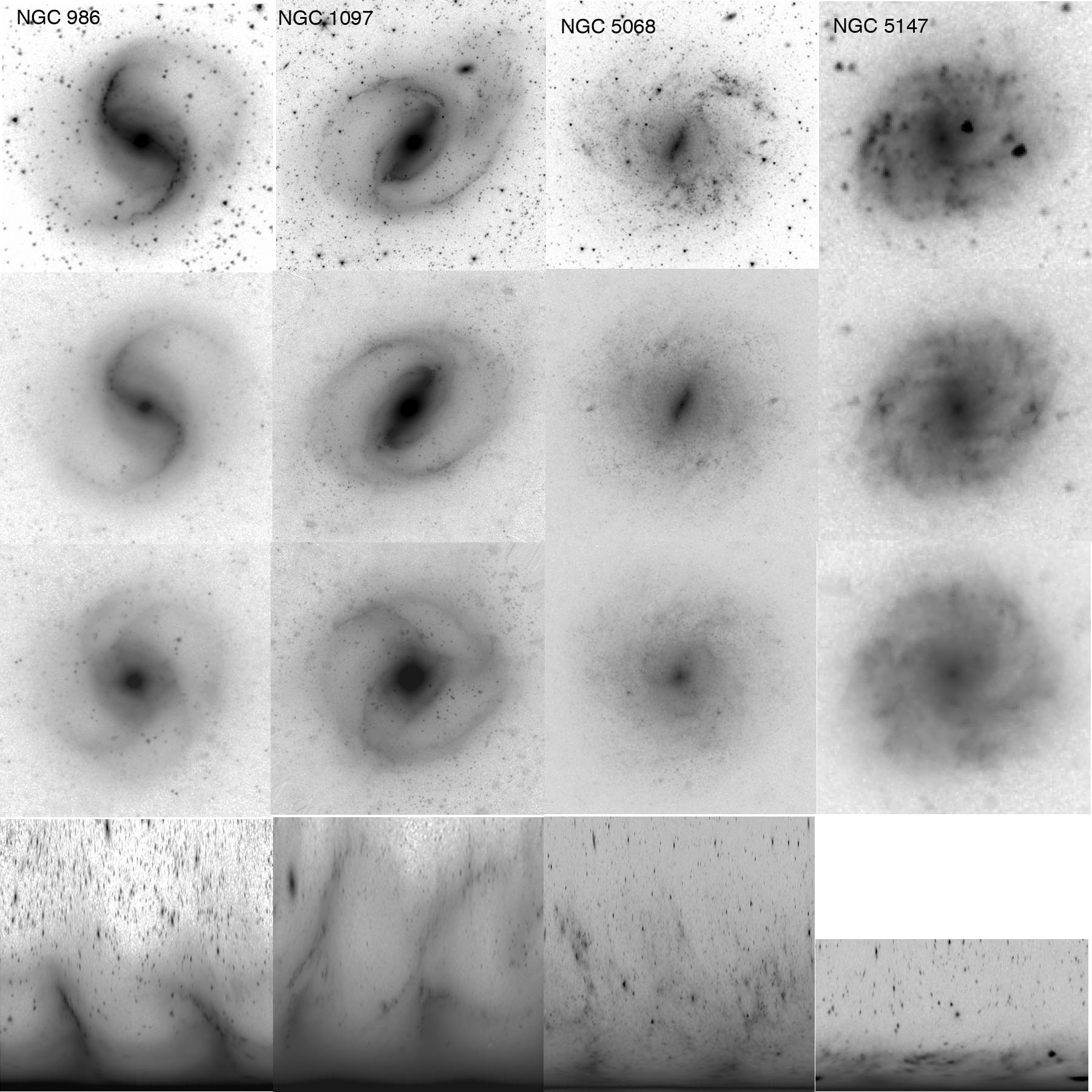}
%\plotone{arm-fig4bar-1-2-11.jpg}
\caption{Examples of barred galaxies; same convention as in Figure \ref{GD}. The left two galaxies, NGC 986 and NGC 1097, are early type grand design galaxies, while the right-hand galaxies, NGC 5068 and NGC 5147, are late type flocculent galaxies. The flocculent barred galaxies have weaker bars and arms than the grand design barred galaxies.}\label{bar}\end{figure}

Non-SB grand design galaxies in our sample often show distinct broad
outer spirals beyond and separate from the main inner bright spirals.
The inner spirals, as in NGC 1566 (Fig. \ref{GD}), are associated with
bright star formation and dustlanes that are probably spiral shocks
(e.g., Roberts 1969). Beyond that there are sometimes separate arms
that are smoother and without concentrated star formation.  In NGC
1566, the outer arms begin on each side of the galaxy at
$\sim0.5R_{25}$, and extend at least as far as the edge of the image at
$\sim1.4R_{25}$, forming a pseudo-ring. Also in NGC 1566, there is a
ridge of star formation disjoint from and leading the main outer spiral
in the south. This ridge is described as a ``plume'' by \cite{buta07}.
Plumes usually occur in barred galaxies where they appear as short
disjoint arms of enhanced emission at the ends of bars in the leading
direction \citep{buta84, buta95}.

Other non-SB grand design galaxies in our survey have disjoint smooth
outer arms too (Fig. \ref{GD}). NGC 4321 has two strong inner arms with
considerable $m=3$ symmetry, and outside of these starting at about
$0.5R_{25}$ is a pair of spirals at higher pitch angle. The outer arms
are clear in the top image of NGC 4321 in Figure \ref{GD}. On the left
of the image there is a branch from the inner spiral to the outer
spiral, which continues toward the companion galaxy NGC 4322. Just as
in NGC 1566, these outer arms are broader and smoother than the inner
arms, they contain little concentrated star formation, and they extend
far out in the image.

NGC 5194 has an interacting companion galaxy that could have modified
any outer spirals, so we do not consider it in this context here.
Still, \cite{tully74} and others noted that the spiral arms in M51 have
a kink in the middle and seem to be composed of two separate arm
systems. We see that kink also at the 7 o'clock position of the bright
outer arm in Figure \ref{GD}.  \cite{dobbs10} reproduced the kink in a
companion-interaction model of M51, and noted that it arose at the
intersection of an old, inner spiral pattern, generated by the first
fly-by of the companion, and a new, outer spiral pattern generated by
the second fly-by.

NGC 5248 in Figure \ref{GD} has strong and symmetric inner arms and
faint, broad, and disjoint outer arms. The outer arms were also present
in a deep $R$-band image in \cite{jog02} and they were first described by
\cite{bur62} in a deep $B$-band image. In the northwest, there is a
second ridge or short arm that is disjoint from the main outer arm, as
in the southern plume of NGC 1566.

The grand design SB galaxies in Figure \ref{bar} do not have disjoint
outer arms. Instead, the inner arms continue smoothly to the outer
galaxy. There is also no obvious counterpart to smooth disjoint outer
arms in the multiple arm galaxies of Figure \ref{MA}. Multiple arm
galaxies generally have an inner 2-arm symmetry, as seen in the figure,
and narrow, irregular and asymmetric arms with star formation all the
way to the edge of the image.

\begin{figure}\epsscale{1}
\plotone{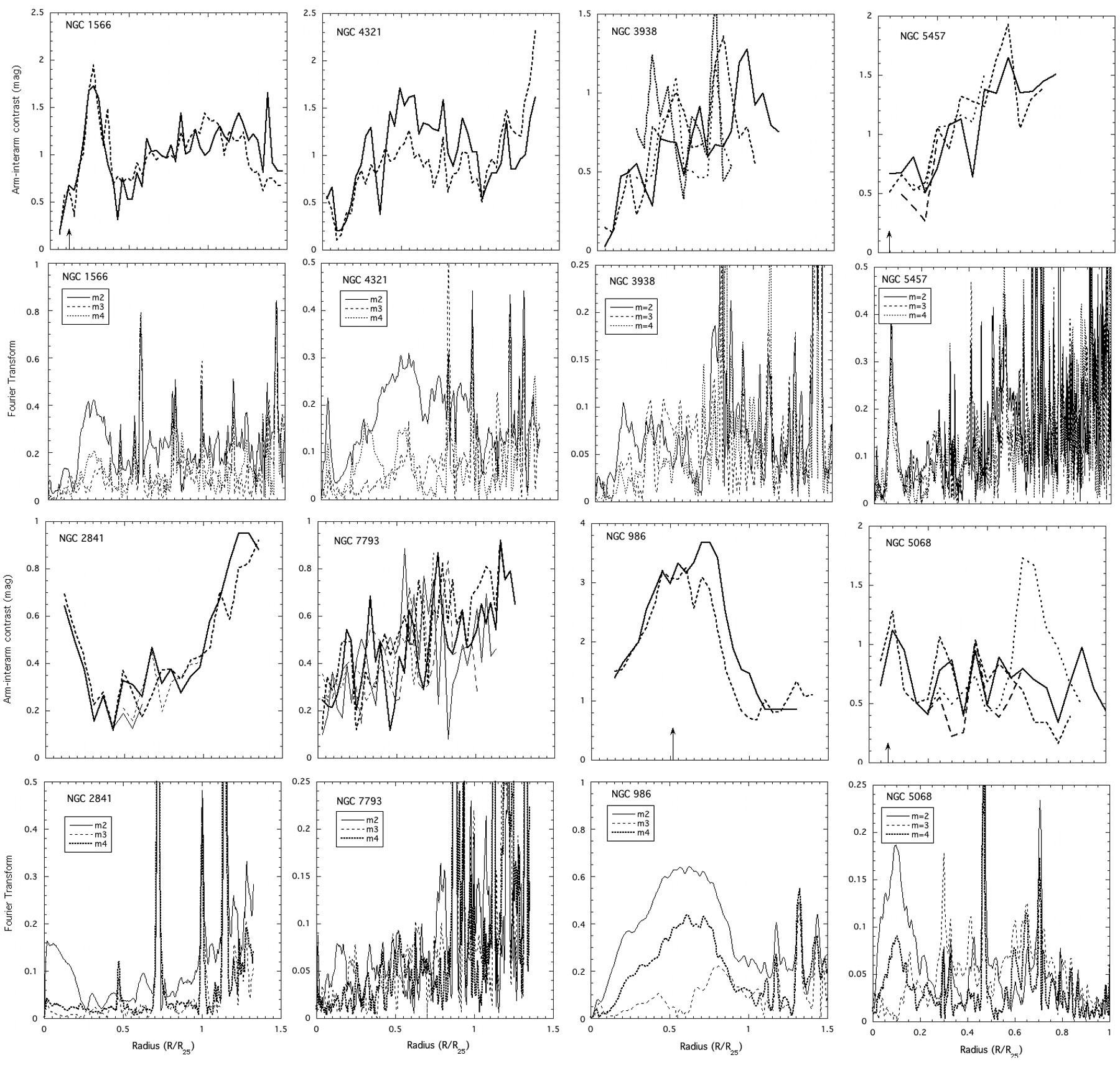}
%\plotone{arm-fig5aiFT-try.jpg}
\caption{Plots of arm-interarm contrasts (1st and 3rd rows) and Fourier
transforms (2nd and 4th rows). The  lines in each arm contrast figure
are the arm-interarm contrasts for each of the two main arms, or the
bar-interbar contrasts in the inner regions of SAB and SB galaxies. The
Fourier transforms are for the m=2, 3, and 4 components. The m=2
component is strongest in each case. The narrow spikes in some Fourier
transforms are due to foreground stars. Enlarge this figure onscreen
for ease in viewing details. }\label{armint}\end{figure}

\begin{figure}\epsscale{1}
\plotone{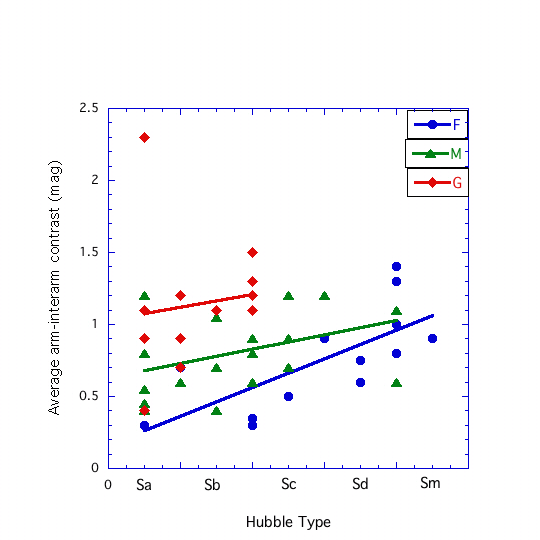}
%\plotone{avgarmintHT,FMG.jpg}
\caption{Average arm-interarm contrast (in magnitudes) is plotted as a function of Hubble type, sorted by Arm Class (blue dots = flocculent, green triangles = multiple arm, red diamonds = grand design). Linear fits for each Arm Class are shown as solid lines. (The high contrast grand design galaxy is NGC 4314; even excluding it, the fit for grand design galaxies lies above the multiple arm and flocculent galaxies). For a given Hubble type, grand design galaxies have stronger arms than flocculent galaxies. Later Hubble types have slightly stronger arms than earlier types. In our sample the later types are multiple arm or flocculent.}\label{armHTAC}\end{figure}

\begin{figure}\epsscale{1}
\plotone{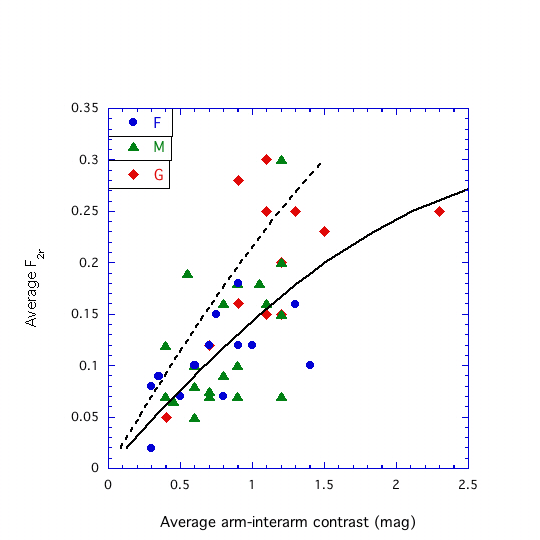}
%\plotone{avgF2rvavgarm,fit.jpg}
\caption{The average over the disk (not including a bar, if present) for the m=2 component  plotted versus the average of the arm-interarm contrast, sorted by Arm Class; grand design galaxies are shown as red diamonds, multiple arm as green triangles, and flocculent as blue dots. The lines show expected correlations as follows: the dashed line is the relation if the arm-interarm intensity is given by (1+2$F_{2r}$)/(1-2$F_{2r}$). The solid line is the fit if the arm intensity includes the m=4 component, taken to be 0.5 $F_{2r}$, so that the intensity is given by (1+3$F_{2r}$)/(1-3$F_{2r}$).  }\label{fit}\end{figure}

\begin{figure}\epsscale{1}
\plotone{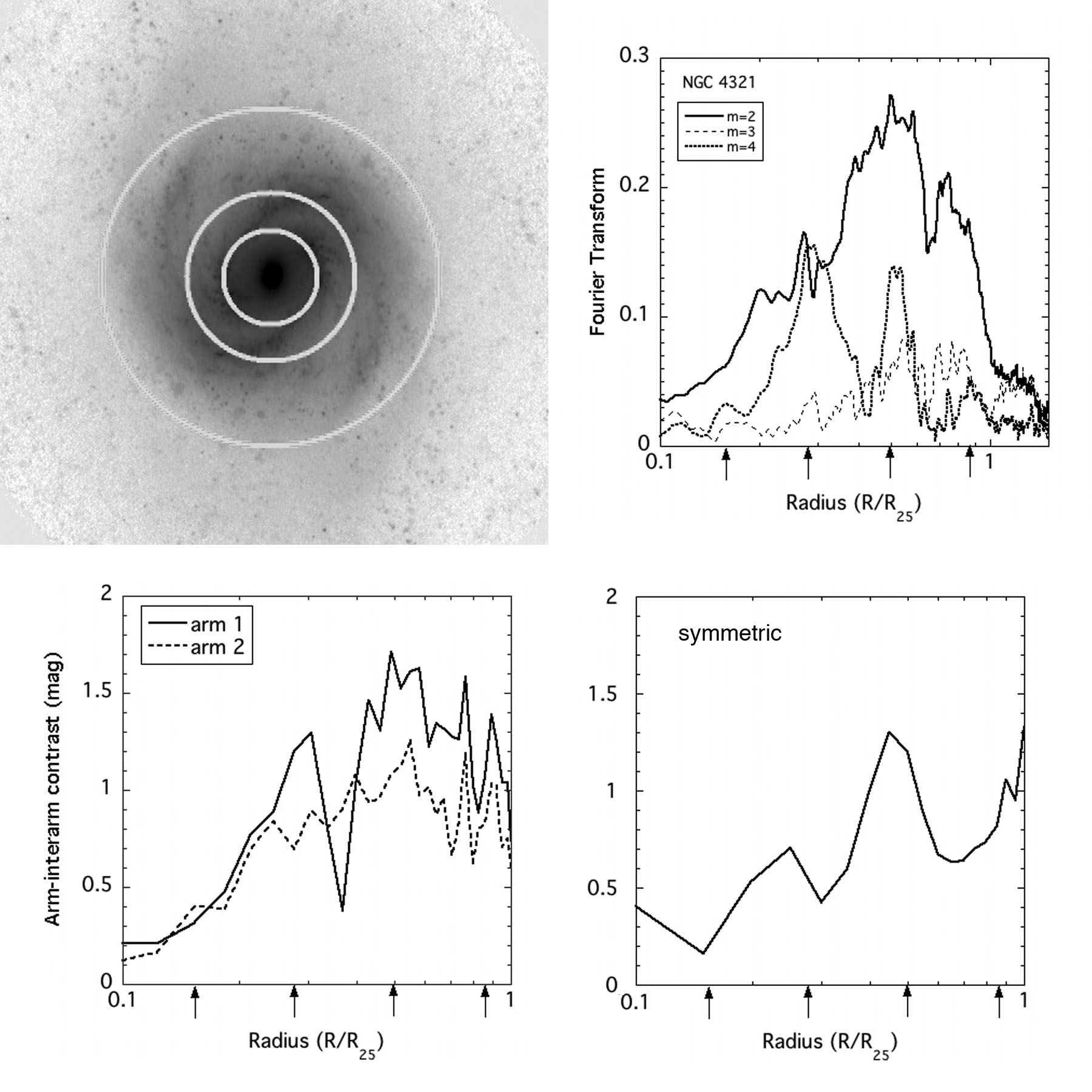}
%\plotone{N4321symarmint.jpg}
\caption{NGC 4321 shows some evidence of spiral arm modulation in the
variation of arm-interarm contrast with radius. (top left) The
deprojected symmetric  image of NGC 4321 is shown with circles overlaid
to indicate where peaks occur in the arm-interarm plots. (top right)
Fourier transforms for the $m=2, 3,$ and $4$ components are shown.
(lower left) Arm-interarm contrasts are shown for each of the two main
arms as a function of radius. The $x$ axis for radius is shown on a
logarithmic scale to emphasize features in the inner regions. (lower
right) Arm-interarm contrasts are shown versus radius based on the
symmetric image shown in the upper right.}\label{sym4321}\end{figure}

Presumably the outer arms in non-SB grand design galaxies are stellar
features that are excited by the inner arms or inner oval. There is no
evidence from physical connections with the inner arms that the outer
arms have the same pattern speed.  The outer arms, if they are density
waves, should extend only to their own outer Lindblad resonance (OLR).
For a flat rotation curve, the OLR is at 1.707 times the corotation
radius. Considering the $m=2$ Fourier transforms in Figure
\ref{armint}, the dip at $0.45R_{25}$ could be corotation, because this
is also where the bright star formation ridge ends. If the pattern
speed for the inner and outer arms were the same, then the OLR would be
at $0.77R_{25}$. This is in the middle of the outer arm, and so
impossible for a standard spiral density wave. Alternatively, if
corotation were at the outer edge of the second peak in the $m=2$
Fourier transform, at $0.8R_{25}$, then the OLR would be at
$1.36R_{25}$ for a flat rotation curve. This is close to the outer edge
of the third peak in the $m=2$ Fourier transform and the end of the
outer spiral.  This is an acceptable fit, but it implies the
unconventional interpretation that corotation is well beyond the end of
the star formation ridge.

A second possibility is that the outer arms are driven by manifolds
emanating from the Lagrangian points at the ends an inner oval disk.
The viewing angles we used to rectify the image were chosen to force
the disk to be circular. Different values are found from kinematics.
\cite{pence90} found a position angle of $41 \pm 3$ degrees and an
inclination of $27 \pm 3$ degrees from their H$\alpha$ Fabry-Perot
data. Deprojecting with these viewing angles, we find that the inner
part stays oval and thus manifold-driven spirals are an alternative.
These can extend well beyond the OLR. Also, the shape of the arms with
the kinematic deprojection is in agreement with a permissible manifold
shape (see e.g. Figures 4 to 6 of Athanassoula et al. 2009a and Sect.
4.3 of Athanassoula et al. 2009b).   Multiple arm or flocculent
galaxies would not have smooth outer arms in the manifold
interpretation because they would not create the strong $m=2$
perturbations required.

A third possibility is that the outer spiral is a resonance response to
the inner spiral at a different pattern speed
\citep{sellwood85,tagger87}. For example, the OLR of the inner spiral
could be the source of excitation at corotation for the outer spiral.
For inner spiral corotation at the edge of the star formation ridge in
NGC 1566, at $0.45R_{25}$, and an OLR of the inner spiral and
corotation of the outer spiral at $0.77R_{25}$ (in agreement with OLR found by Elmegreen \& Elmegreen 1990) for a flat rotation
curve, then the OLR of the outer spiral would be at $1.3R_{25}$. This
is also an acceptable fit because that is about the radius where the
outer spiral ends.

In NGC 4321, the ridge of star formation in the south ends at about
$0.4R_{25}$ in the north and beyond that there is a second arm with a
small pitch angle extending to the north until $\sim0.7R_{25}$.  A
similar ridge and second arm is on the other side of the galaxy. These
features make the two peaks in the $m=2$ Fourier transform plot of
Figure \ref{armint}. The second arm could lie between corotation and
the OLR of the inner spiral. The outer spiral mentioned above is beyond
that, from $\sim R_{25}$ to $\sim1.3R_{25}$, as shown by a broad ledge
in the $m=2$ Fourier transform plot. As for NGC 1566, the outer spiral
extends too far to end at the OLR of the inner spiral if the inner
spiral has corotation at the end of a star formation ridge.  The outer
spiral in NGC 4321 could be a manifold or resonance phenomenon too.
With corotation of the inner spiral at $0.4R_{25}$, and corotation of
the outer spiral at the OLR of the inner spiral, at $0.7R_{25}$, the
OLR of the outer spiral would be at $1.2R_{25}$ for a flat rotation
curve. This is about the extent of the outer spiral.

These examples suggest that smooth outer spirals observed at $3.6\mu$m
in non-SB grand design galaxies are either driven by manifolds
surrounding an oval inner disk, or excited by a resonance with the main
inner spirals. The resonance may be one where corotation of the outer
spiral is at the OLR of the inner spiral. Barred (SB) galaxies may have
similar manifolds and resonance excitations but in that case most of
the main spiral would be in this form \citep{sellwood88,lia09a}, and no
additional smooth spirals would exist beyond that, unless they are
higher-order excitations and too faint to see here. Multiple arm and
flocculent galaxies, which do not show smooth outer spirals in this
survey, could lack sufficiently strong $m=2$ perturbations in the inner
disk to drive them.

\subsection{Sharp Edges in Spiral Arms}
\label{sect:sharp}

The outer southern spiral in NGC 4321 has what appears to be a sharp
edge in Figure \ref{GD}; that is, an abrupt transition from the outer
arm to the disk or sky. The western arm in NGC 5194 and the inner arms
in NGC 1566 also have sharp outer edges. These outer edges do not look
as sharp on azimuthal profiles because they are stretched out by a
factor equal to the inverse tangent of the pitch angle.

\begin{figure}\epsscale{1}
\plotone{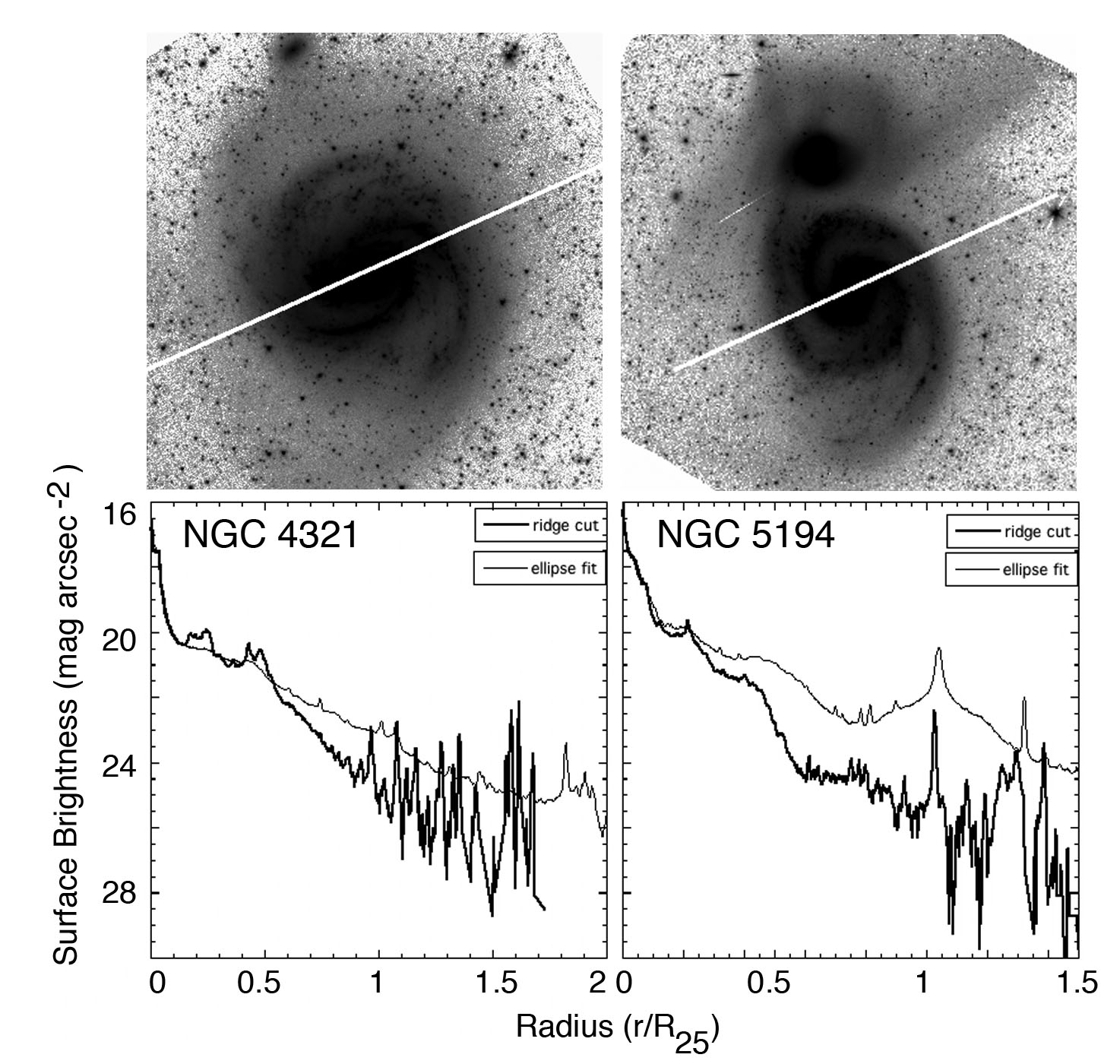}
%\plotone{fig-Ridge.jpg}
\caption{(top) Logarithmic intensity images in channel 1 are shown for NGC 4321 (left) and NGC 5194 (right), with a line indicating where the 10-pixel-wide cut was made for a radial profile. (bottom) Radial profiles for the right-hand-side of the 10-px cuts are shown as heavy lines, and azimuthally-averaged profiles from ellipse fits are shown as thin lines. Note the steep fall-off in the profiles past the arms.}\label{ridge}\end{figure}

The intensities along strips cutting nearly perpendicular to the spiral
arms and going through the galaxy centers are shown for NGC 4321 and
NGC 5194 in Figure \ref{ridge}. The strips are 10 pixels wide to reduce
noise. Sharp drop-offs occur at the outer parts of several spiral arms.
In both galaxies, the arms at $\sim0.5R_{25}$ show a steep decline on
the outer edge. Linear least-square fits to these drop-offs reveal that
they have an approximately exponential profile about twice as steep as
the local disk measured on ellipse-fit profiles, which averages over
azimuth.

A sharp outer edge corresponds to a strong amplitude spiral superposed
on an exponential disk.  A typical azimuthal profile of a grand design
galaxy (EE84) has a sinusoidal variation in surface brightness, which
is a logarithmic intensity scale. Assume the amplitude of this
variation is $\mu_0$. Considering also the exponential disk with scale
length $r_D$, this spiral profile means that the intensity varies with
radius $r$ and azimuthal angle $\theta$ approximately as
\begin{equation}
I(r,\theta)=I_0\exp\left(-r/r_D+0.4(\ln10)\mu_0\sin(2[\theta(r)-\theta_0])\right),
\end{equation}
where $\mu_0$ is the surface brightness in units of mag arcsec$^{-2}$.
For a logarithmic spiral,
\begin{equation}
\theta(r)=\theta_0+\ln(r/r_0)/\tan(i)
\end{equation}
with spiral pitch angle $i$. The inverse of the local scale length at
the outer part of the arm, where the radial gradient is largest, is
given by
\begin{equation}
{{dI}\over {Idr}} = -{1\over {r_D}} - {{0.8(\ln 10})\mu_0 \over {r\tan
i}}.
\end{equation} The first term is from the underlying disk and the
second term is from the outer part of the spiral arm. Evaluating the
second term, $0.8(\ln10)=1.84$,  $r\sim2r_D$ for these spirals, and $\tan
i\sim0.27$ for $i=15^\circ$. Thus the second term is approximately
$3\mu_0/r_D$ for $\mu_0\sim1$ in magnitudes. It follows that the total
gradient can be $\sim4$ times the underlying disk gradient for a strong
spiral.

Grand design spiral arms can have sharp outer edges that are comparable
in scale to the epicyclic radius for stars. This is consistent with
theoretical predictions that the arms are non-linear waves in which a
large fraction of the stars have their epicycles in phase in spiral
coordinates. Sharp edges also imply strong radial force gradients,
softened by the disk thickness, which is comparable to the scale length
there.  Such radial forcing causes the arm amplitudes to grow by
locking in more and more stars to the common epicycle pattern.

Sharp outer edges in tidally-induced arms are present in recent
simulations by \cite{oh10}.  The sharp edges result from a growing
accumulation of stars in a moving wave front at the galaxy edge. They
are a caustic in the distribution of particle orbits \citep{struck90,
elmegreen91}. The sharp edge in the southwest of M51 is also in the
simulation by \cite{dobbs10}.

\subsection{Barred galaxies}
\label{sect:bar} Our sample includes 18 SB, 18 SAB, and 10 SA galaxies.
The average value of the Fourier transform for the $m=2$ component in
the arms is weaker in non-barred galaxies than in barred galaxies; the
averages are $0.15\pm0.042$, $0.26\pm0.12$, and $0.34\pm0.14$ for SA,
SAB, and SB galaxies, respectively. Similarly, the arm-interarm
contrast is weaker in non-barred galaxies, with averages $0.69\pm0.030$,
$0.94\pm0.33$, and $0.94\pm0.44$ for SA, SAB, and SB galaxies,
respectively. These results indicate that the presence of bars or ovals
increases the amplitudes of the arms.

\begin{figure}\epsscale{1}
\plotone{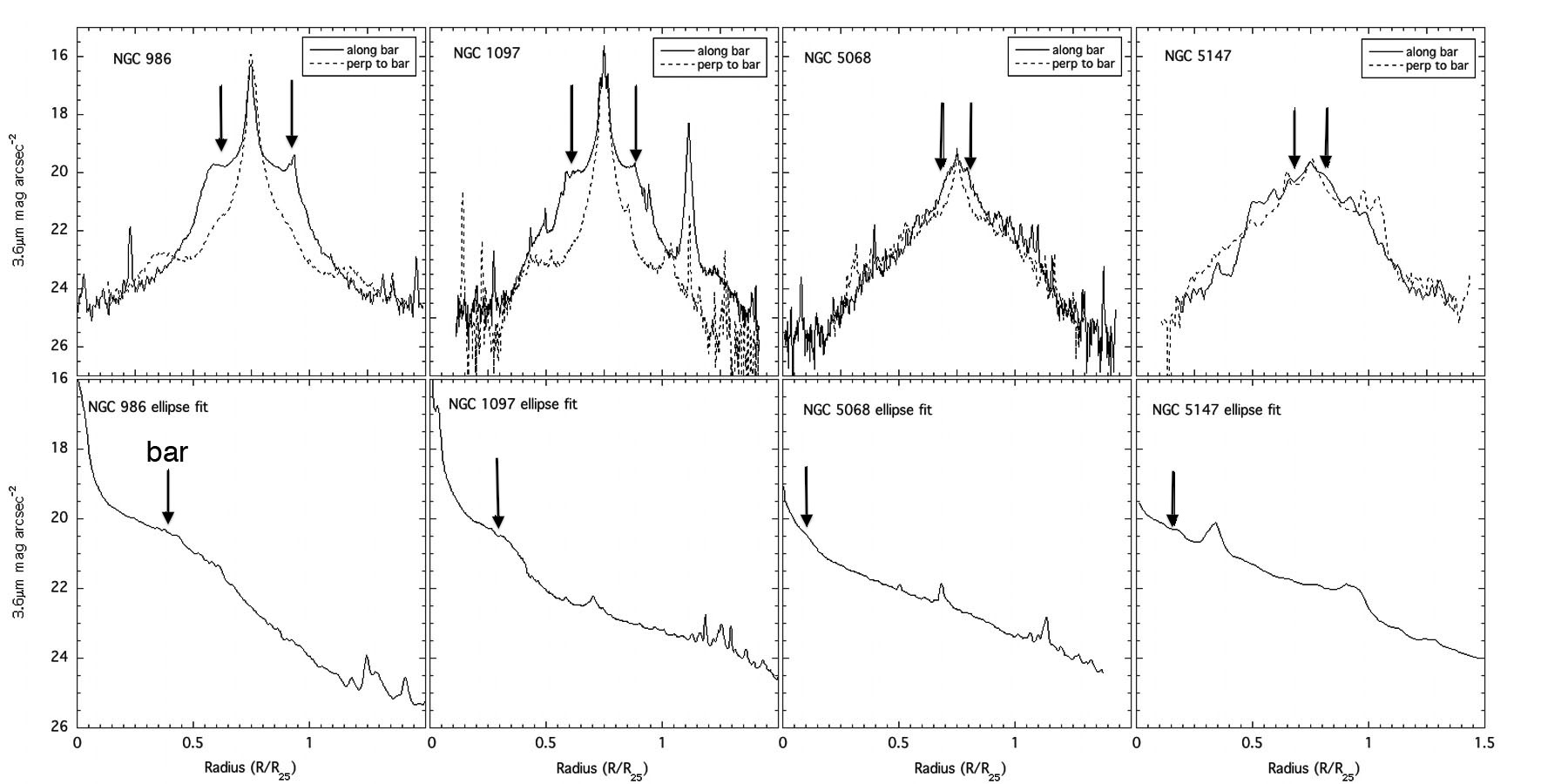}
%\plotone{arm-fig6-barprof.jpg}
\caption{Radial profiles are shown for the galaxies in Figure \ref{bar}. The top row shows cuts along the bars and perpendicular to the bars; the bottom row shows azimuthally averaged radial profiles based on ellipse fits. NGC 986 and NGC 1097, early type galaxies, have ``flat'' bars, in which the surface brightness has a slower decline than the exponential disk; NGC 5068 and NGC 5147, late type galaxies, have  ``exponential'' bars, in which the bar is nearly indistinguishable from the underlying disk.}\label{barcont}\end{figure}

Figure \ref{barcont} shows bar intensity profiles for the four galaxies
in Figure \ref{bar}. Two of these galaxies are grand design early types
(NGC 986 and NGC 1097) and two are flocculent late types (NGC 5068 and
NGC 5147). For all barred galaxies in the sample, profiles were
measured along and perpendicular to the bar (shown in the top row of the figure for these 4 galaxies), and from azimuthal
averages based on $ellipse$ fits (shown in the bottom row in the figure). Image counts were converted to
surface brightness using a formula in the online IRAC Instrument
Handbook: $\mu$(AB mag arcsec$^{-2})=20.472 -
2.5\times\log$(Intensity[MJy/sr]). The figure shows a difference in the
profiles for the early and late types. The bars in early type galaxies are long and
their major axis profiles are flat (that is, slowly declining), as are their azimuthally averaged
profiles. The late type bars are short and have an exponential decline similar to the disk, with just a small flattening on
the major axis. Their azimuthal profiles hardly show the bars at all. Evidently, bars in early type galaxies reorganize the inner disk regions of their
galaxies, but bars in late type galaxies do this to a much lesser extent.

Flat or exponential bar profiles
are listed in Table \ref{table2}.  In our whole sample, the SB galaxies
are dominated by flat bars (13 flat, 5 exponential), while the SAB
galaxies have mostly exponential bars (3 flat, 15 exponential). Among
the early type galaxies, SAB and SB galaxies are somewhat biased
toward flat bars (15 flat, 10 exponential), while the late type galaxies are
dominated by exponential bars (1 flat, 10 exponential). In our sample, all of the grand
design barred galaxies have flat bars, while all of the flocculent barred
galaxies have exponential bars.

\begin{figure}\epsscale{1}
\plotone{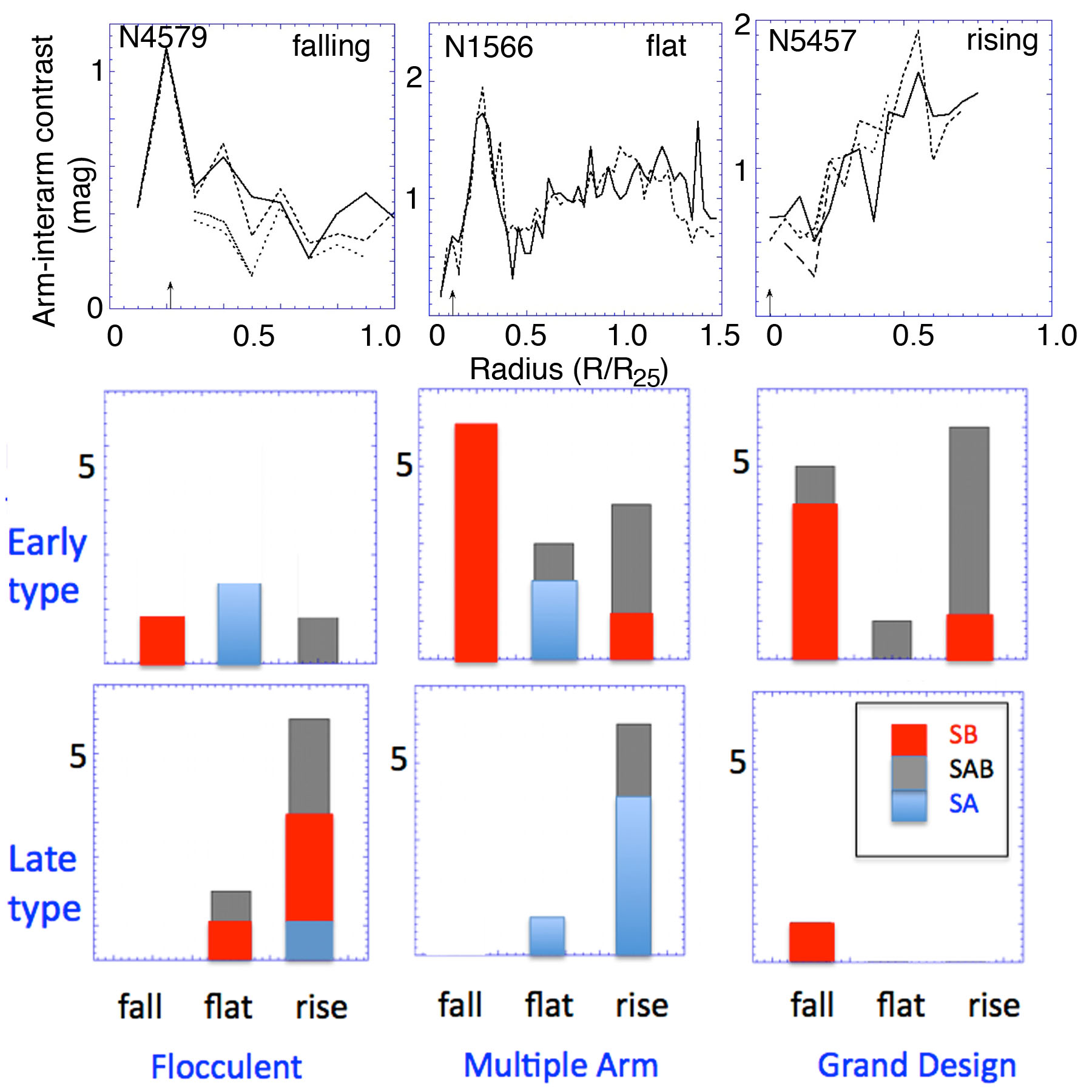}
%\plotone{slopearm,type,ac.jpg}
\caption{(top) Arm-interarm contrast as a function of radius (in units of $R_{25}$) for 3 galaxies, NGC 4579, NGC 1566, and NGC 5457. The arrows along the abscissa indicate the end of a bar or oval. The arms beyond that have contrasts that fall, remain constant, or rise, respectively. (middle, bottom) Histograms are shown for early type (middle) and late type (bottom) for falling, flat, or rising arm-interarm contrasts, for flocculent, multiple arm, and grand design galaxies, subdivided by SA (blue), SAB (gray), and SB (red) galaxies. SB grand design and multiple arm tend to be falling, while SA and SAB of all Arm Classes are rising or flat. Flocculent and late types tend to be rising.}\label{slope}\end{figure}

Arm-interarm contrasts in the spiral region vary as a function of
radius (Fig. \ref{armint}), with either falling, constant, or rising
trends. These trends are listed in the last column of Table
\ref{table3}. Three examples of radial contrast variations are shown in
Figure \ref{slope}: falling in NGC 4579, constant in NGC 1566, and
rising in NGC 5457. The arrows on the abscissa indicate the locations
of the ends of the bars or ovals. The middle and bottom panels of
Figure \ref{slope} show histograms for the early and late types, coded
for falling, flat, or rising arm-interarm contrasts, and subdivided
according to Arm Class. The galaxies are also coded by bar types SA,
SAB, and SB.  The SB grand design and multiple
arm galaxies in our sample tend to have falling arm contrasts with radius,
while the SA and SAB galaxies for all Arm Classes tend to have rising or flat
arm contrasts in our sample. Flocculent and late type galaxies tend to
have rising arm contrasts. We also find that 11 out of 12 galaxies with
falling arm-interarm contrasts have flat bars (10 of which are SB and
all of which are multiple arm or grand design), while only 3 out of 18
galaxies with rising arm-interarm contrasts have flat bars (and all of
these are multiple arm or grand design).

Falling spiral arm amplitudes beyond SAB and SB bars could be an
indication that these bars end near corotation \citep{contop}.
Corotation is generally where spiral waves are amplified, in both the
WASER theory \citep{lau76} and the swing amplifier theory
\citep{toomre81}. The arm amplitudes decrease away from corotation and
generally approach zero amplitude at the inner and outer Lindblad
resonances, where the waves are absorbed. Thus, the decreasing
amplitudes can be an indication that the bars are driving the spirals
where these two features meet, and that they both have the same pattern
speed.  The decrease in arm amplitude beyond the bars is also
consistent with predictions by the ``manifold theory'' of spiral
structure \citep{romgoma,romgomb,lia09a,lia09b,athan10}. For strong
bars, the manifolds have the shape of spiral arms and widen with
increasing radius, thereby diminishing the arm-interarm contrast. If
there is interference with other spiral components, e.g. with a weak
leading component, this could lead to bumps on an otherwise smoothly
decreasing density profile.

\begin{figure}\epsscale{1}
\plotone{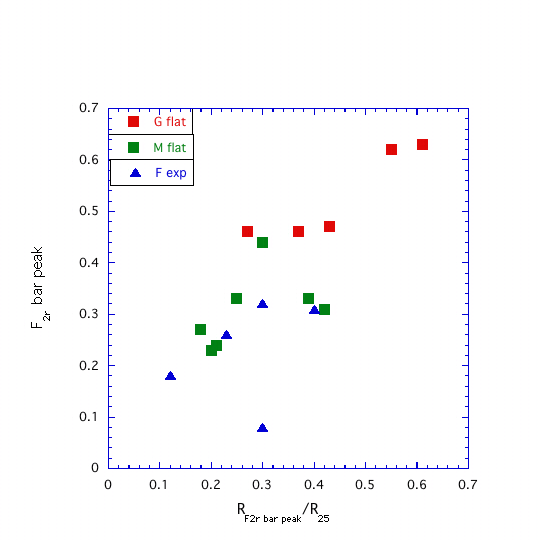}
%\plotone{F2rvsR2rSB.jpg}
\caption{Peak of the Fourier transform for m=2 in the bar for SB galaxies as a function of normalized radial distance of the peak, sorted by Arm Class (blue = flocculent, green = multiple arm, red = grand design) and bar profile (square = flat type, diamond = exponential type). Stronger bars are longer, which was seen in previous optical and $K_s$-band observations also. Bars are progressively stronger from flocculent to multiple arm to grand design galaxies. All of the flat barred SB galaxies are multiple arm or grand design, while all of the exponential barred SB galaxies are flocculent.}\label{baramp}\end{figure}

The peak value of $F_{2}$ in the bar is a measure of bar amplitude.
Table \ref{table3} lists this peak value along with the radius (in
units of $R_{25}$) at which it occurs. The radius of the peak is slightly less than, but correlated with, the end of the bar, according to simulations \citep{athan02}.  In Figure \ref{baramp} the peak
$F_2$ amplitude in the bar for SB galaxies is shown as a function of
radius at which this peak occurs. Longer bars tend to have higher peak
$F_2$ amplitudes, as found also in optical and near-IR work mentioned
previously and by \cite{EE07}. The Spearman's rank correlation
coefficient is 0.68, with a significance at the 99\% confidence level.
A bivariate least-squares fit to the relation gives a slope of
1.16$\pm$0.328 for peak versus relative bar length.

This result is consistent with ideas of secular bar evolution
\citep[e.g.,][]{debat00,athan02,athan03,val03, athan09}. In simulations
of bar evolution \citep{athan03}, angular momentum is emitted mainly by
near-resonant material in the bar and absorbed mainly by near-resonant
material in the outer disk and halo. As bars lose angular momentum,
they can become more massive and/or thinner and/or longer. In the two
first cases, the peak $m=2$ amplitude increases, while in the third one
the radius of the peak amplitude increases.

Our Fourier transform results are similar to those found in studies of
barred galaxies in other passbands. Two of the galaxies in $K_s$-band
studies by \cite{EE07}, NGC 986 and NGC 7552, are in our current
sample. The $m=2$ Fourier component of NGC 986 in $K_s$ band has a peak
of 0.62 at a radius of $0.6R_{25}$, and NGC 7552 has a peak value of
0.6 at a normalized radius of 0.6;  we find the same peaks and peak
radii in $3.6\mu$m to within $\sim10$\%, as shown in Table
\ref{table2}.  EE85 have NGC 3504, NGC 4314, and NGC 7479 in common
with our list. Their $I$-band $m=2$ bar components are 0.52, 0.8, and
0.8, compared with our $3.6\mu$m values of 0.53, 0.45, and 0.45; NGC
3504 is about the same in $I$ band and $3.6\mu$m, but the other two are
stronger in $I$ band.

\begin{figure}\epsscale{1}
\plotone{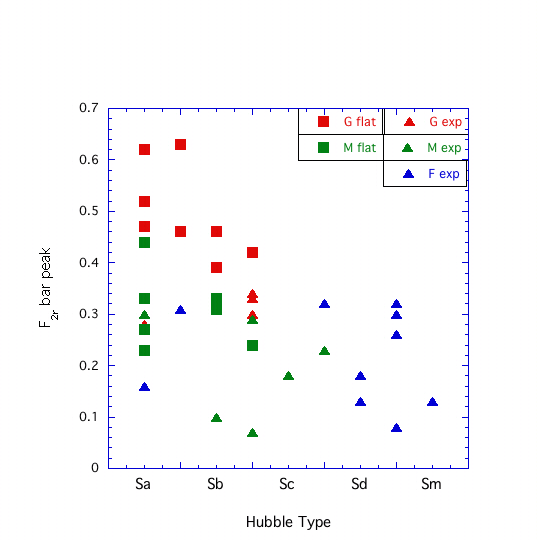}
%\plotone{figF2rHT.jpg}
\caption{Peak of the Fourier transform for m=2 in the bar for SAB and SB galaxies as a function of Hubble type, sorted by Arm Class (blue = flocculent, green = multiple arm, red = grand design) and bar profile (square = flat type, diamond = exponential type). The strongest bars are in early types with flat bars and grand designs, while the weakest bars are in later types with exponential bars and flocculent structure. Multiple arms are a mix of bar strengths and bar profiles.}\label{barHT}\end{figure}

Figure \ref{barHT} shows the $F_{2}$ bar peak as a function of Hubble
type for SAB and SB galaxies, sorted by Arm Class. There is a steady
decrease of bar amplitude with later Hubble types, although the
correlation is weak; the Pearson correlation coefficient is 0.403,
significant at the 98\% confidence level. Early types tend to be grand
design (for the highest amplitude bars) or multiple arm (for weaker
amplitude bars), while later types tend to be flocculent with weak
bars. The early types are predominantly flat bars, while the later
types are mostly exponential bars. These results are consistent with
optical (EE85) and near-IR \citep{regan97} studies. The exponential
SABs have a mixture of Arm Classes, while the exponential SBs are all
flocculent and mostly late type. Conversely, most of the flocculent
exponential-barred galaxies are late-type, while most of the multiple
arm and all of the grand design exponential-barred galaxies are early
type. All of the flat bars of either SAB or SB type are early Hubble
type with multiple arm or grand design spiral structure. These results
reinforce the idea that high amplitude flat bars of either SAB or SB
type occur in early Hubble types and drive spirals in the outer disks.

There are no grand design non-barred galaxies in our sample. Of the 10
SA types, 3 are flocculent and 7 are multiple arm. Among the early type
SA galaxies, 2 are flocculent and 2 are multiple arm, while among the
late type SA galaxies, 1 is flocculent and 5 are multiple arm. This
observation reinforces the idea that bars or oval distortions help
drive grand design spirals. Of course, our sample is small; there could
be non-barred isolated grand design galaxies that were not studied in
this paper. (M81 is an example of a non-barred grand design galaxy,
although it has interacting companions.) \cite{kend11} studied a SINGS
sample of 31 galaxies; 15 of their 17 barred galaxies (SAB or SB) have
multiple arm or grand design structure, while only 7 of their 13
non-barred (SA) galaxies do. When their sample is divided into early
and late type galaxies, the 11 early type SAB or SB galaxies include 1
flocculent, 5 multiple arm, and 5 grand design galaxies, i.e., early
type bars are strongly correlated with stellar spirals. Their 6 late
types include 1 flocculent, 5 multiple arm, and no grand designs, which
means late type bars are correlated with weak or no stellar spirals.

\begin{figure}\epsscale{1}
\plotone{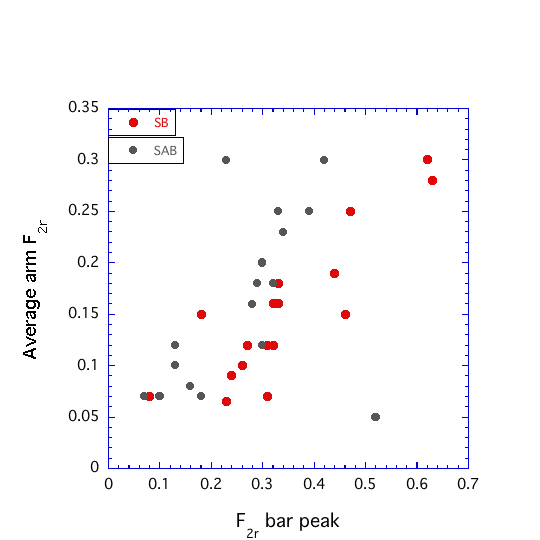}
%\plotone{avgF2armvsF2bar.jpg}
\caption{Average of the Fourier transform for m=2 in the arm for SAB and SB galaxies as a function of Fourier peak m=2 for the bar, sorted by bar type (gray = SAB, red = SB). The bar amplitudes are correlated with average arm amplitudes, supporting the idea that bars drive waves.}\label{bararm}\end{figure}

Figure \ref{bararm} shows the average $F_{2}$ spiral arm amplitude
versus the peak $F_{2}$ bar amplitude, sorted by SAB and SB types.
There is a weak correlation with larger amplitude bars corresponding to
larger amplitude arms. The Pearson correlation coefficient is 0.59 for
the plotted points, increasing to 0.85 if only SB galaxies are
considered, with a significant at the 99\% confidence level. A
bivariate least-squares fit to the points gives a slope of
0.49$\pm$0.08. This correlation has been seen in the near-IR also
\citep{buta09,salo10}. The result suggests that bars drive stellar
waves, and that stronger bars drive stronger waves.

There is little systematic difference in the bar amplitudes between SAB
and SB bar types. In the figure, the SB types are slightly shifted to
the lower right, suggesting slightly larger amplitudes for SB compared
with SAB bars for a given arm amplitude. However, it is not generally
true that SB types are strong bars and SAB types are weak bars. Both
types have a range of bar amplitudes that correlate in the same way
with Hubble type (Figure \ref{barHT}).  The SAB types have lower
eccentricities than the SB types, with about the same amplitudes.

\cite{seigar1} and \cite{seigar03} found no connection between bar and
arm strengths in their $K_s$-band study of 45 spirals, although they
did not plot Fourier transform results. Instead, they used a measure of
arm strength they called the ``equivalent angle,'' based on measuring
the angle subtended by a disk segment containing the same flux as the
arm or bar at a given radial range. Their Figure 8 shows a wide
variation of arm strength for a given bar strength, but generally
smaller bar strengths correspond with smaller arm strengths.

\section{Conclusions}
\label{sect:conc} Deep $Spitzer$ images of spiral galaxies at $3.6\mu$m
allow measurements of arm structure out to $\sim1.5R_{25}$, further out
than many previous optical and near-IR studies. We selected a
representative sample of 46 galaxies to measure underlying old stellar
disks in a variety of spiral Arm Classes, Hubble types, and bar types.
The morphology does not change much from optical to mid-IR, although
some optically flocculent galaxies show subtle 2-arm structure at
$3.6\mu$m.

The early types in our sample tend to have multiple arm and grand
design spirals, while the late types have flocculent spirals. Barred
early types also tend to have high amplitude bars that are relatively
long and flat in radial profiles, and outside of these bars are
generally grand design or multiple arm spirals that decrease in
amplitude with radius. Barred late type galaxies typically have low
amplitudes and short exponential bars with flocculent disks. Bars with
higher amplitudes are correlated with higher average arm amplitudes.
These trends are consistent with results from optical data, and support
theories of secular bar growth and the driving of grand design spiral
structure by strong bars.

The outer spiral arms of non-barred grand design galaxies appear to be
disjoint from and smoother than the inner arms. They do not have the
obvious ridges of star formation that also characterize inner disk
arms, but there can be plumes of star formation nearby. Barred grand
design galaxies do not have such disjoint arms, nor do multiple arm or
flocculent galaxies. Most likely, the spiral arms in barred galaxies
with strong bars extend between corotation near the end of the bar and
the outer Lindblad resonance; they may have different pattern speeds,
as discussed by many authors. The outer spirals in non-barred grand
design galaxies could also extend to the outer Lindblad resonance at
the same pattern speed as the inner spirals, or they could be a
resonance feature at a different pattern speed.

%A study of the larger sample of spiral galaxies in the full S$^4$G study will provide a good statistical sample to confirm these results.

{\it Acknowledgments:} We are grateful to the dedicated staff at the
Spitzer Science Center for their help and support in planning and
execution of this Exploration Science program. We also gratefully
acknowledge support from NASA JPL/Spitzer grantRSA1374189 provided for
the S$^4$G project. DME gratefully acknowledges Vassar College for
student support for Andrew Yau, and thanks NASA/JPL/Caltech for grant
1368024.  EA and AB thank the Centre National d'Etudes Spatiales for
financial support. This research is based in part on archival data
obtained with the Spitzer Space Telescope, which is operated by the Jet
Propulsion Laboratory, California Institute of Technology under a
contract with NASA, and has made use of the NASA/IPAC Extragalactic
Database (NED) which is operated by the Jet Propulsion Laboratory,
California Institute of Technology, under contract with the National
Aeronautics and Space Administration. The National Radio Astronomy
Observatory is a facility of the National Science Foundation operated
under cooperative agreement by Associated Universities, Inc. We thank
the referee for helpful comments on statistics.

\clearpage

\begin{deluxetable}{lcccccc}
\tabletypesize{\scriptsize}\tablecolumns{7} \tablewidth{0pt} \tablecaption{Galaxy properties}

\tablehead{\colhead{NGC}& \colhead{Arm Class\tablenotemark{a}}&\colhead {Bar type\tablenotemark{b}}&\colhead{mid-IR type\tablenotemark{b}} & \colhead{PA\tablenotemark{c}} & \colhead{inclination\tablenotemark{d}}&\colhead{ Radius\tablenotemark{d}}  \\
\colhead{}&\colhead{}&\colhead{}&\colhead{}&\colhead{(degrees)}&\colhead{(degrees)}&\colhead{(arcmin)}}
\startdata
300 &   M   &   A   &   dm  &   107.7   &   44.9    &   10.94   \\
337 &   F   &   AB  &   cd  &   130 &   50.9    &   1.44    \\
428 &   F   &   AB  &   dm  &   114.3   &   40.7    &   2.04    \\
628 &   M   &   A   &   c   &   107.5   &   24.2    &   5.24    \\
986 &   G   &   B   &   ab  &   123.2   &   40.7    &   1.95    \\
1097    &   G   &   B   &   ab  &   126.9   &   47.5    &   4.67    \\
1313    &   F   &   B   &   dm  &   16.2    &   40.7    &   4.56    \\
1433    &   M   &   B   &   a   &   15.6    &   24.2    &   3.23    \\
1512    &   M   &   B   &   a   &   67.7    &   50.9    &   4.46    \\
1566    &   G   &   AB  &   b   &   21.1    &   37.4    &   4.16    \\
2500    &   F   &   AB  &   d   &   71.4    &   24.2    &   1.44    \\
2552    &   F   &   AB  &   m   &   53.7    &   48.7    &   1.73    \\
2805    &   M   &   AB  &   c   &   182.4   &   40.7    &   3.16    \\
2841    &   F   &   AB  &   a   &   147.4   &   64.1    &   4.06    \\
2903    &   M   &   B   &   b   &   23.4    &   61.4    &   6.30    \\
3049    &   G   &   B   &   ab  &   26.6    &   48.7    &   1.09    \\
3147    &   M   &   AB  &   b   &   111.7   &   27.0    &   1.95    \\
3184    &   M   &   A   &   bc  &   118.2   &   21.1    &   3.71    \\
3198    &   M   &   AB  &   bc  &   34.9    &   67.1    &   4.26    \\
3344    &   M   &   AB  &   bc  &   68  &   24.2    &   3.54    \\
3351    &   M   &   B   &   a   &   20.1    &   47.5    &   3.71    \\
3504    &   G   &   AB  &   a   &   0   &   39.1    &   1.35    \\
3627    &   M   &   B   &   b   &   174.5   &   62.8    &   4.56    \\
3906    &   F   &   B   &   dm  &   0   &   27.0    &   0.93    \\
3938    &   M   &   A   &   c   &   0   &   24.2    &   2.69    \\
3953    &   M   &   B   &   bc  &   15.1    &   59.9    &   3.46    \\
4254    &   M   &   A   &   c   &   66  &   29.4    &   2.69    \\
4299    &   M   &   A   &   dm  &   0   &   21.1    &   0.87    \\
4314    &   G   &   B   &   a   &   55.5    &   27.0    &   2.09    \\
4321    &   G   &   AB  &   bc  &   189.3   &   31.7    &   3.71    \\
4450    &   M   &   A   &   ab  &   172.5   &   42.2    &   2.62    \\
4536    &   G   &   AB  &   bc  &   124.1   &   64.8    &   3.79    \\
4579    &   M   &   B   &   a   &   87.1    &   37.4    &   2.95    \\
4689    &   F   &   A   &   bc  &   73  &   35.6    &   2.13    \\
4725    &   M   &   AB  &   a   &   37.7    &   44.9    &   5.36    \\
4736    &   G   &   AB  &   a   &   130.6   &   35.6    &   5.61    \\
5055    &   F   &   A   &   bc  &   104.7   &   54.9    &   6.30    \\
5068    &   F   &   B   &   d   &   0   &   29.4    &   3.62    \\
5147    &   F   &   B   &   dm  &   112 &   35.6    &   0.95    \\
5194    &   G   &   AB  &   bc  &   105 &   51.9    &   5.61    \\
5248    &   G   &   AB  &   bc  &   129.6   &   43.6    &   3.08    \\
5457    &   M   &   AB  &   cd  &   0   &   0.0 &   0.00    \\
5713    &   F   &   B   &   ab  &   13.5    &   27.0    &   1.38    \\
7479    &   G   &   B   &   b   &   44.5    &   40.7    &   2.04    \\
7552    &   G   &   B   &   a   &   178 &   37.4    &   1.69    \\
7793    &   F   &   A   &   c   &   99.9    &   47.5    &   4.67    \\
\label{table1}
\enddata
\tablenotetext{a}{from \cite{EE87} except for NGC 3906, classified here}
\tablenotetext{b}{from \cite{buta10} except for NGC 4450, 4689, and 5247, from RC3}
\tablenotetext{c}{from mid-IR isophotes of inner disk}
\tablenotetext{d}{from the blue light diameter $D_{25}$ in RC3 }

\end{deluxetable}

\clearpage
\begin{deluxetable}{lcccccc}
\tabletypesize{\scriptsize} \tablewidth{0pt} \tablecaption{Bar and Arm Measurements}
\tablehead{\colhead{NGC}& \colhead{$R_{F2r}/R_{25}$\tablenotemark{a}}&\colhead {$F_{2r}$\tablenotemark{b}}&{bar\tablenotemark{c}}&\colhead{$F_{2r}$ arm\tablenotemark{d}}& \colhead{Arm-interarm\tablenotemark{e}} &\colhead{arm slope\tablenotemark{f}}  \\
\colhead{}& \colhead{}&\colhead {bar peak}&{profile}&\colhead{average} & \colhead{average (mag)} &\colhead{}  }
\startdata
300 &   0.31    &   0.13    &-& 0.08    &   0.6 &   flat    \\
337 &   0.40    &   0.32    &exp&   0.18    &   0.9 &   rise    \\
428 &   0.70    &   0.30    &exp&   0.12    &   1.0 &   rise    \\
628 &   0.45    &   0.20    &-& 0.1 &   0.9 &   rise    \\
986 &   0.61    &   0.63    &flat&  0.28    &   0.9 &   fall    \\
1097    &   0.37    &   0.46    &flat&  0.15    &   1.2 &   fall    \\
1313    &   0.30    &   0.32    &exp&   0.16    &   1.3 &   rise    \\
1433    &   0.30    &   0.44    &flat&  0.19    &   0.55    &   fall    \\
1512    &   0.39    &   0.33    &flat&  0.16    &   0.8 &   fall    \\
1566    &   0.28    &   0.39    &flat&  0.25    &   1.1 &   flat    \\
2500    &   0.18    &   0.13    &exp&   0.1 &   0.6 &   flat    \\
2552    &   0.28    &   0.13    &exp&   0.12    &   0.9 &   rise    \\
2805    &   0.24    &   0.18    &exp&   0.07    &   1.2 &   rise    \\
2841    &   0.07    &   0.16    &exp&   0.08    &   0.3 &   rise    \\
2903    &   0.42    &   0.31    &flat&  0.07    &   0.7 &   fall    \\
3049    &   0.37    &   0.32    &-& 0.12    &   0.7 &   fall    \\
3147    &   0.39    &   0.10    &exp&   0.07    &   0.4 &   flat    \\
3184    &   0.31    &   0.19    &-& 0.1 &   0.6 &   flat    \\
3198    &   0.59    &   0.29    &exp&   0.18    &   0.9 &   rise    \\
3344    &   0.07    &   0.07    &exp&   0.07    &   0.9 &   rise    \\
3351    &   0.18    &   0.27    &flat&  0.12    &   0.4 &   fall    \\
3504    &   0.40    &   0.52    &flat&  0.05    &   0.4 &   fall    \\
3627    &   0.25    &   0.33    &flat&  0.18    &   1.05    &   fall    \\
3906    &   0.23    &   0.26    &exp&   0.1 &   1.4 &   rise    \\
3938    &   0.20    &   0.11    &-& 0.075   &   0.7 &   rise    \\
3953    &   0.21    &   0.24    &flat&  0.09    &   0.8 &   rise    \\
4254    &   0.40    &   0.18    &-& 0.15    &   1.2 &   rise    \\
4299    &   0.52    &   0.17    &-& 0.16    &   1.1 &   rise    \\
4314    &   0.43    &   0.47    &flat&  0.25    &   2.3 &   rise    \\
4321    &   0.55    &   0.30    &exp&   0.2 &   1.2 &   rise    \\
4450    &   0.28    &   0.19    &-& 0.05    &   0.6 &   flat    \\
4536    &   0.30    &   0.33    &exp&   0.25    &   1.3 &   rise    \\
4579    &   0.20    &   0.23    &flat&  0.065   &   0.45    &   fall    \\
4689    &   0.11    &   0.07    &-& 0.02    &   0.3 &   flat    \\
4725    &   0.42    &   0.30    &exp&   0.2 &   1.2 &   rise    \\
4736    &   0.80    &   0.28    &exp&   0.16    &   0.9 &   rise    \\
5055    &   0.54    &   0.13    &-& 0.09    &   0.35    &   flat    \\
5068    &   0.12    &   0.18    &exp&   0.15    &   0.75    &   flat    \\
5147    &   0.30    &   0.08    &exp&   0.07    &   0.8 &   rise    \\
5194    &   0.38    &   0.34    &exp&   0.23    &   1.5 &   rise    \\
5248    &   0.53    &   0.42    &flat&  0.3 &   1.1 &   rise    \\
5457    &   0.56    &   0.23    &exp&   0.3 &   1.2 &   rise    \\
5713    &   0.40    &   0.31    &exp&   0.12    &   0.7 &   fall    \\
7479    &   0.27    &   0.46    &flat&  0.15    &   1.1 &   fall    \\
7552    &   0.55    &   0.62    &flat&  0.3 &   1.1 &   fall    \\
7793    &   0.65    &   0.16    &-& 0.07    &   0.5 &   rise    \\

\label{table2}
\enddata
\tablenotetext{a}{ratio of radius of peak m=2 Fourier transform in bar to blue light $R_{25}$ radius. Uncertainties in the m=2 components are $\sim\pm0.02$}
\tablenotetext{b}{bar peak Fourier transform for m=2}
\tablenotetext{c}{flat or exponential bar, determined from radial profiles}
\tablenotetext{d}{arm peak Fourier transform for m=2}
\tablenotetext{e}{average arm-interarm contrast in disk}
\tablenotetext{f}{slope of arms (beyond bar or oval, if present)}

\end{deluxetable}
\clearpage
\begin{deluxetable}{lccccc}
\tabletypesize{\scriptsize} \tablewidth{0pt} \tablecaption{Arm Comparisons by Band}
\tablehead{\colhead{NGC}& \colhead{B-band\tablenotemark{a}}&\colhead {I-band\tablenotemark{a}}&\colhead{$K_s$-band\tablenotemark{b}}&\colhead{3.6$\mu$m} & \colhead{4.5$\mu$m}   \\
\colhead{}& \colhead{(mag)}&\colhead {(mag)}&\colhead{(mag)}&\colhead{(mag)} & \colhead{(mag)}   }
\startdata
628 &0.86&  0.67    &&  0.95&0.70   \\
2500&0.9&0.6&&0.6&\\
2841&   0.48    &   0.19&   &   0.3 &   0.3     \\
2903&0.76&0.48&(1.3)&0.9,(1.2)&\\
3344&   0.86    &   0.86    &&  0.8 &       \\
3504&0.9&0.5&&0.5&\\
3627&&&1.8&1.5&\\
4254    &   1.1 &   0.9 &   &1.4    &       \\
4314&0.8&1.0&2.2&2.0&\\
4321&0.95&0.67&&1.3&1.3\\
5055&0.38&0.28&&0.40&\\
5194&1.6&1.7&&1.3&\\
5248&1.5&1.8&&1.1&\\
5457&0.95&0.86&&1.0&\\
7479&1.2&1.4&&1.0&\\
7793    &   0.76    &   0.29    &&  0.45    &   0.45    \\

\label{table3}
\enddata
\tablenotetext{a}{ Arm-interarm contrasts were evaluated at approximately the same radius in all bands for a given galaxy. Uncertainties are $\sim$0.1 mag; from \cite{EE84}}
\tablenotetext{b}{from \cite{regan97}}
\end{deluxetable}

\clearpage
%fig1

\clearpage
 %fig2

\clearpage
 %newfig3

\clearpage
 %fig4

\clearpage
%fig5

\clearpage
%fig6

\clearpage
% fig7

\clearpage
% fig8

\clearpage
%fig9

\clearpage
 %fig10

\clearpage
 %fig11

\clearpage
 %fig12

\clearpage
 %fig13

\clearpage
 %fig14

\end{document}